\documentclass[aps,showpacs,preprintnumbers,longbibliography]{revtex4-2}

\usepackage{amsmath,amssymb}
\usepackage{graphicx}
\usepackage{natbib}
\usepackage{siunitx}
\usepackage{xcolor}
\usepackage{epstopdf,textcomp} 

\newcommand{\bnabla}{\boldsymbol{\nabla}\:}

\begin{document}

\title{Electroosmotic flow in small-scale channels induced by surface-acoustic waves}

\author{Mathias Dietzel} \email{mdietzel.edu@e.mail.de} \author{Steffen Hardt} \email{hardt@nmf.tu-darmstadt.de}
\affiliation{Technische Universit\"at Darmstadt, Fachbereich Maschinenbau, Fachgebiet Nano- und Mikrofluidik\\
Alarich-Weiss-Strasse 10, D-64287 Darmstadt, Germany}

\date{\today}

\begin{abstract}
In this work, numerical simulations of the Navier-Stokes, Nernst-Planck, and the Poisson equations are employed to describe the transport processes in an aqueous electrolyte in a parallel-plate nanochannel, where surface-acoustic waves (SAWs) are standing or traveling along (piezo-active) channel walls. It is found that -in addition to the conventional acoustic streaming flow- a time-averaged electroosmotic flow is induced. Employing the stream function-vorticity formulation, it is shown that the Maxwell stress term causes an electroosmotic propulsion that is qualitatively identical to the one discussed in the context of alternating current (AC) electroosmosis (EOF). Differences arise mainly due to the high actuation frequencies of SAWs, which are in the MHz range rather than in the kHz-regime typical for ACEOF. Moreover, the instantaneous spatial periodicity of the EOF in the travel direction of the SAW is intrinsically linked to the dispersion relation of the latter rather than a free geometric parameter. This leads to a specific frequency band where an EOF of sizable magnitude can be found. On the low frequency end, the ratio between the electric double layer (EDL) thickness and the SAW wavelength becomes extremely small so that the net force leading to a non-vanishing time-averaged EOF becomes equally small. On the high frequency end, the RC time of the EDL is much larger than the inverse of the SAW frequency. This leads to both a vanishing effective charge density of the EDL as well as to inertial effects at the boundary between the EDL and the bulk of the electrolyte. For a parallel-plate channel, the EOF can be maximized by using two SAWs on both channel walls that have the same frequency but are phase-shifted by $180^\circ$. This maximizes the electric field and the corresponding electric current across the channel that drives the mechanical imbalance of the ion cloud in the EDL. For instance, at $10$ MHz, $9\:\textrm{nm}$ (nominal) EDL thickness, and a plate separation of $480\:\textrm{nm}$, a (time-averaged) plug-like EOF with peak velocity of ${\cal O}(10^{-1})\:\textrm{mm s}^{-1}$ is found. Although reliable information about the acoustic streaming velocity under such confinement is scarce, it appears that the SAW-EOF is larger than that and therefore the dominant pumping mechanism for such a scenario. The proposed actuation might be a viable alternative for driving liquid electrolytes through narrow ducts and channels, without the need for electric interconnects and electrodes. \\ \\
This is the post-print authors' version of the manuscript, which was published in Physical Review Fluids. doi:10.1103/PhysRevFluids.5.123702  \copyright \ American Physical Society 2020.
\end{abstract}


\maketitle

\section{Introduction} \label{Sec:intro}

Surface-acoustic waves (SAWs) are sound waves traveling along solid surfaces with a speed of $c_\textrm{SAW} = \lambda_\textrm{SAW} f_\textrm{SAW}$ of the order of $3500\:\textrm{ms}^{-1} < c_\textrm{SAW} < 4000\:\textrm{ms}^{-1}$ at frequencies of $10\:\textrm{kHz} < f_\textrm{SAW} < 100\:\textrm{MHz}$ and wavelengths of the order of $35\:\mu\textrm{m} < \lambda_\textrm{SAW} < 0.4\:\textrm{m}$. Most commonly, they are generated with interdigitated transducer (IDT) electrode combs, which are fabricated lithographically on a piezo-active substrate such as a single crystal lithium niobate (LiNbO$_3$, LN) wafer \cite{Franke:LabChip2009} and actuated by a periodic voltage signal. The comb spacing is chosen such that it approximates $\lambda_\textrm{SAW}$. Applying electric power to the IDT leads to the emission of Rayleigh waves traveling along the substrate surface and away from the IDT. If such a wave reaches an area where the substrate is in contact with a liquid, the SAW leaks into the liquid at a refraction (Rayleigh) angle determined by $\theta_\textrm{R} = \textrm{sin}^{-1}(c_\textrm{liq}/c_\textrm{SAW})$. For common liquids, the speed of sound is given by $c_\textrm{liq} \approx 1450\:\textrm{ms}^{-1}$, so that $\theta_\textrm{R} \approx 22\:^{\circ}$ {\cite{Dentry:PRE2014}. The transition into the liquid to form a bulk acoustic wave (BAW) typically proceeds on three length scales relative to the viscous boundary (Stokes) layer thickness $\delta_\eta = \sqrt{2 \eta/(\rho \omega_\textrm{SAW})}$ \cite{Wiklund:LabChip2012} as well as to $\lambda_\textrm{SAW}$. In the definition of $\delta_\eta$, $\eta$ and $\rho$ denote the dynamic viscosity and the density of the liquid, respectively, while the angular frequency is given by $\omega_\textrm{SAW} = 2 \pi f_\textrm{SAW}$. For water and the frequency range given above, one finds $50\:\textrm{nm} \leq \delta_\eta \leq 5\:\mu\textrm{m}$. For length scales $l \leq \delta_\eta$, the leaking sound wave drives the (inner) viscous boundary layer or Schlichting streaming, which is a steady rotational fluid motion. For $\delta_\eta < l < \lambda_\textrm{SAW}$, the Schlichting vortices are balanced by counter-rotating vortices in the outer boundary layer, commonly referred to as Rayleigh streaming \cite{Wiklund:LabChip2012,Sadhal:LabChip2012,Wu:Fluids2018}. Finally, for $l > \lambda_\textrm{SAW}$ the attenutation of the sound wave leads to a gradient in an acoustic radiation pressure that is the source of Eckart or "quartz wind" streaming \cite{Eckart:PhysRev1947,Riley:AnnRevFluidMech2001}. As a rough classification, Rayleigh-Schlichting or boundary-driven streaming \cite{Nyborg:JAcoustSocAm1958} occurs in the near field while Eckart streaming is a far-field phenomenon. Since the acoustic actuation is sinusoidal in time, for a fully linear system no time-averaged velocity would be observed. In the simplest case, the propagation of the sound pressure wave is described by a linear viscous wave equation whose solution includes a coefficient defined by Stokes' law of sound attenuation in a Newtonian fluid \cite{Stokes:TCPS1845}. This viscous attenuation, caused by the dissipation-induced lag between pressure actuation and flow reaction, gives rise to a non-linear body force in the fluid and to the observation of a finite net flow velocity \cite{Riley:AnnRevFluidMech2001,Wiklund:LabChip2012}. Actuation by SAWs has received vivid attention for spraying applications \cite{Kurosawa:SensAct1995,Qi:PoF2008} and inducing a fluid motion in small liquid entities such as drops and films \cite{Franke:LabChip2009,Yeo:AnnRevFluidMech2014}. Corresponding flow velocities largely depend on geometric factors but may reach ${\cal O}(10^1)\:\textrm{cm s}^{-1}$ \cite{Beyssen:SensActB2006,Yeo:Biomicrofluidics2009,Dentry:PRE2014}. Such high velocities cannot be induced in strongly confined domains such as microchannels, since for channel widths smaller than $\lambda_\textrm{SAW}$ or even smaller than $\delta_\eta$ the conventional Rayleigh-Schlichting streaming is disturbed. While it has been shown that adding a wall texture whose amplitude is of similar order as $\delta_\eta$ may improve streaming \cite{Lei:MANO2018}, even for relatively wide channels with a free (no-stress) surface on one side, flow velocities do not exceed ${\cal O}(10^1)\:\textrm{mm s}^{-1}$ \cite{Tan:EPL2009}. Since the pressure wave in the liquid propagates at an angle of $\theta_\textrm{R}$ compared to the SAW, it is continuously reflected (and refracted) at the channel walls if the channel is wide enough, with only the reflected and subsequently interfering BAWs leading to a net flow in the direction along the channel. Studies of SAW-driven liquid transport in channels bounded by solid walls focus mostly on cases where the SAW propagation is perpendicular to the channel main axis. For instance, circular flow patterns with velocities of ${\cal O}(10^0-10^1)\:\textrm{mm s}^{-1}$ can be generated \cite{Schmid:MANO2012,Kiebert:LabChip2017}, and Hags\"ater \textit{et al.} observed flow velocities of the order of a few tens of $\mu \textrm{m s}^{-1}$ in a microfluidic chamber \cite{Hagsaeter:LabChip2007}. In addition, SAW-induced acoustic streaming may cause liquid atomization and subsequent drop coalescence in a channel, causing filling velocities of ${\cal O}(10^0)\:\textrm{mm s}^{-1}$ in a direction opposing the SAW propagation \cite{Cecchini:APL2008}. To date, estimates of transport velocities in nanochannels driven by acoustic effects are solely simulation-based. Here, a clear judgment is difficult as the velocities are typically reported in a dimensionless fashion. For instance, Xie and Cao report by means of molecular dynamics (MD) simulation "fast nanofluidics" in nanochannels driven by SAWs \cite{Xie:MANO2017}, but converting their findings into physical units by employing their non-dimensionalization would imply transport velocities of more than $100\:\:\textrm{m s}^{-1}$. Tan and Yeo mention at one point of their numerical study based on the Lattice-Boltzmann approach that the average velocities do not exceed ${\cal O}(10^{-5})\:\textrm{m s}^{-1}$ \cite{Tan:PRF2018}, which appears to be a more realistic estimate.

In a piezo-active material, the displacement of the substrate atoms due to the propagating SAW wave is proportional to the (complex) electrostatic surface potential given by $U_\textrm{SAW} = \hat{U}_\textrm{SAW}\exp[i(\omega_\textrm{SAW}t-\boldsymbol{k}_\textrm{SAW} \cdot \boldsymbol{x})]$ \cite{Jakubik:SensActB2014}, where $\hat{U}_\textrm{SAW}$, $\boldsymbol{k}_\textrm{SAW}$, and $\boldsymbol{x} = (x,y,z)^\textrm{T}$ denote the amplitude of the surface potential, the SAW wave vector, and the position vector, respectively. In this work, it is assumed that the SAW substrate is in contact with a dilute aqueous electrolyte, i.e., with a (moderately) conducting liquid, so that the flow actuation principle discussed herein would be weakened by a charge transfer between both. Yet, it has recently been shown that acoustic streaming induced by SAWs can still be present in strong electrolytes \cite{Huang:AdvMat2020}, while LN acoustic plate sensors may detect liquid electrolytes based on the conductivity-dependent frequency shift and attenuation of the SAW \cite{Josse:SensActB1992}. The latter finding was mainly attributed to the polarization of the electrolyte by the electric field via migration of dissolved ions rather than to the reduction of the surface potential by conduction through the electrolyte between oppositely charged areas on the LN substrate. In addition, the performance of LN as an anode material in contact with aqueous electrolytes has been studied using galvanostatic charge-discharge measurements \cite{Son:ElectrochemComm2004} and cyclic voltammetry \cite{Lui:MatLett2014}. In these experiments, a pronounced capacitive behavior of LN was proven. In turn, this indicates that an electric double layer (EDL) builds up on the surface of the material. This claim is supported by an experimental study in which a liquid micro-lens array is generated on top of a LN substrate by electrowetting effects and driven by the pyroelectric behavior of LN \cite{Grilli:OpticsExpr2008}. Together, these studies suggest that typical setups for (small-scale) acoustofluidics can be used to address SAW-driven electroosmotic flow (EOF) as well. Considering the high SAW frequencies, the ion cloud in the vicinity of the wall cannot reach a state of mechanical equilibrium, or in other terms, the Gibbs-Duhem equation, expressing that the sum of the gradients in chemical potential (at constant temperature) of each electrolyte component equals zero, is not fulfilled \cite{Fitts:McGrawHill1962}. This leads to an electroosmotic propulsion (EOP), which originates from the SAW along the surface and adds to the flow induced by acoustic streaming. Typical SAW flow domains are large in comparison to the EDL thickness so that electric fields and affiliated currents in between oppositely charged regions, ultimately driving the EOP, are small. In the case that narrow channels are considered, where both channels walls are subjected to a SAW, these currents and the corresponding EOP can be large, leading to flow that might exceed the one induced by acoustic effects. This is the focus of the present paper.

The underlying physical principle of SAW-induced EOP is identical to the one of induced-charge (IC) EOF \cite{Squires:JFM2004} and alternating current (AC) EOF. In the simplest form of EOF, an externally applied (i.e., source-free) electric field acts on the charge density of an EDL formed at a charged wall. Over the past two decades, a plethora of technological applications for such electrokinetic flows has emerged. For instance, one can take advantage of the large surface-to-volume ratio to perform sensing applications in micro-total-analysis systems ($\mu$-TAS) \cite{Suh:Micromachines2010,Luka:Sensors2015,Olanrewaju:LabChip2018}, enhance the cooling of micro-processors \cite{Murshed:RenewSustainErgRev2017}, or to embody small-scale energy conversion systems \cite{Yang:JMicroMechEng2003,vdHeyden:PRL2005}, in which mechanical energy is partially converted into electric energy. For moderate external fields, the ion cloud is in a state of mechanical equilibrium, implying that the osmotic and the electrostatic forces caused by the local electric field within the EDL exactly cancel each other. Most prominently, this is the case for solutions that invoke the classical Poisson-Boltzmann (PB) theory. Nevertheless, the mechanical balance can be disturbed -for instance- by applying gradients in bulk ion concentration or temperature, giving rise to diffusoosmosis \cite{Liu:Langmuir2013,Jing:PCCP2018} or thermoosmosis \cite{Bregulla:PRL2016,Dietzel:JFM2017}, respectively. Furthermore, using permselective membranes as wall material permits electric currents passing through them, which -beyond certain thresholds of the driving electric field- may lead to concentration polarization and a mechanical imbalance of the ion cloud in the vicinity of the wall. As a consequence, an electrohydrodynamic (EHD) instability develops, which is the reason for the experimentally observed overlimiting current through nanopores \cite{Rubinstein:PRE2000}. By the same principle, the application of large or time-varying electric fields at the walls as used in ICEOF and ACEOF, respectively, may disturb the mechanical equilibrium of the ion cloud near them as well.

For the latter, the equations governing the ion distribution and the electric potential remain non-linear even in the limit of the Reynolds number $Re \rightarrow 0$, so that the time-averaged ACEOF does not vanish \cite{Green:PRE2000,Gonzalez:PRE2000}. In the thin EDL limit and disregarding the Stern layer, the flow velocity at the surface can be approximated by $\langle u \rangle= \epsilon/(4 \eta) \bnabla_\textrm{s}[(\Delta U)^2]$ \cite{Gonzalez:PRE2000}. The dielectric permittivity of the electrolyte solution is denoted by $\epsilon$, while the spatial gradient along the electrode surface is given by $\bnabla_\textrm{s}$. The voltage drop across the diffuse part of the EDL is denoted by $\Delta U$. Classically, two electrodes separated by a small gap are arranged along the flow boundary to apply peak-to-peak AC potentials of ${\cal O}(10^0)\textrm{V}$, leading to time-averaged vortex patterns with peak velocities of ${\cal O}(10^{-1} - 10^{0})\:\textrm{mm s}^{-1}$ at ${\cal O}(10^0)\:\textrm{kHz}$ \cite{Green:PRE2000}. The characteristic time scale is given by the RC time of the EDL defined by $t_\textrm{RC} = (\lambda_\textrm{D} h)/D$ \cite{Squires:JFM2004}, where $h$ and $D$ denote the channel width and the ion diffusivity, respectively. The nominal EDL thickness is given by $\lambda_\textrm{D} = \sqrt{\epsilon k_\textrm{B} T/(2 e^2 \nu^2 n_0)}$, with $k_\textrm{B}$, $T$, $e$, $\nu$, and $n_0$, denoting the Boltzmann constant, the absolute reference temperature, the elementary charge, the valence of the symmetric $\nu:\nu$ electrolyte, and the reference ion concentration in the bulk, respectively. Using much lower frequencies than $t^{-1}_\textrm{RC}$ provides sufficient time for the ions to attain a mechanical equilibrium, while significantly exceeding $t^{-1}_\textrm{RC}$ leads to a practically uncharged EDL, so that both limits go along with vanishing ACEOF. Using an array of electrodes placed along the flow boundaries together with appropriate phase differences of the driving voltages according to a traveling AC wave may lead to unidirectional flow \cite{Ramos:JAP2005,Ramos:JCollIntScie2007,Gonzalez:MANO2008,Yang:IEEETransDielectricsElectricIns2009,Yeh:PRE2011,Hrdlicka:Electrophoresis2014}. The magnitude of the corresponding flow largely depends on the ability to pattern the flow boundary with narrow-spaced and interconnected electrodes.

The work reported in this paper can be viewed from two different perspectives. First, it is a fundamental study of the electrokinetic effects induced by SAWs in narrow channels. In the second perspective, the idea of traveling-wave (TW) ACEOF is picked up, but surface acoustic waves (SAWs) moving or standing along a (piezo-active) flow boundary are considered to induce the alternating electric wall potential instead of micro-fabricated electrode arrays. In the following, the mathematical description used to compute the flow field is detailed, succeeded by the numerical implementation and its verification by means of known results from ACEOF. Subsequently, parameter variations are performed and discussed.

\section{Mathematical description} \label{Sec:math}

The momentum equation for an incompressible, (Newtonian) liquid electrolyte is given by
\begin{equation} 
\label{Eq:NSE} \rho[\partial_t \boldsymbol{u} + (\boldsymbol{u}\cdot\bnabla)\boldsymbol{u}] = -\bnabla p +\eta \bnabla^2 \boldsymbol{u} - \rho_\textrm{f} \bnabla{\phi},
\end{equation}
where the partial time derivative, the velocity vector, and the fluid pressure are denoted by $\partial_t \equiv \partial/\partial t$, $\boldsymbol{u} = (u,v,w)^\textrm{T}$, and $p$, respectively. The last term on the right-hand side (RHS) of Eq. (\ref{Eq:NSE}) describes the Maxwell stress, where the charge density is given by $\rho_\textrm{f}$, while $\phi$ denotes the electric potential. The latter two are connected via the Poisson equation according to
\begin{equation} 
\label{Eq:Poisson} \bnabla^2 \phi = -\frac{\rho_\textrm{f}}{\epsilon}.
\end{equation}
In this work, the frequency dependence of $\epsilon$ is neglected since for water this becomes significant only for frequencies above $1\:\textrm{GHz}$ \cite{Kaatze:JChemEngData1989}. The charge density can be expressed by the number concentrations of the ion species. For a symmetric electrolyte, one finds
\begin{equation} 
\label{Eq:charge_dens} \rho_\textrm{f} = e \nu (n_+ - n_-).
\end{equation}
In turn, the ion number concentrations $n_\pm$ are determined by the Nernst-Planck equations (NPEs) given by
\begin{equation} 
\label{Eq:NPE} \partial_t n_\pm + \boldsymbol{u} \cdot \bnabla n_\pm = \bnabla \cdot [D_\pm \bnabla n_\pm + e \nu_\pm \mu_\pm n_\pm \bnabla \phi],
\end{equation}
where the electrophoretic mobilities of the cation (subscript "+") and the anion (subscript "-") are given by the Stokes-Einstein relation according to $\mu_\pm = D_\pm/(k_\textrm{B} T)$, while the valences are given by $\nu_+=-\nu_-=\nu$. To focus on the essential physics, constant and equal diffusion coefficients for both ion species are assumed, i.e., $D_\pm = D$.

This work focuses on flows through planar, parallel-plate channels so that a description in two spatial dimensions is sufficient. For such flows, the stream function-vorticity formulation can be employed. Following the standard procedure, the pressure can be removed from Eq. (\ref{Eq:NSE}) to read
\begin{equation} 
\label{Eq:NSE_stream} \rho[\partial_t \omega + \partial_y \psi \partial_x \omega - \partial_x \psi \partial_y \omega]\!=\!\eta\!\bnabla^2 \omega + \partial_x \rho_\textrm{f} \partial_y \phi - \partial_y \rho_\textrm{f} \partial_x \phi,
\end{equation}
where $\psi$ denotes the stream function with $\partial_y \psi \equiv \partial \psi/\partial y = u$ and $\partial_x \psi \equiv \partial \psi/\partial x = -v$, while $\omega = \bnabla^2 \psi$. For cases that can be described by the PB theory, the ions in the vicinity of the wall follow the Boltzmann distribution. Consequently, for a symmetric electrolyte with $n_0$ denoting a reference ion concentration at $\phi=0$, the corresponding charge density is given by $\rho_\textrm{f,PB} = -2 e \nu n_0 \textrm{sinh}[e \nu \phi/(k_\textrm{B} T)]$. Thus, in this case at constant temperature and if in addition no external electric field is applied, the electroosmotic propulsion force defined by 
\begin{equation} 
\label{Eq:EOP} F_\textrm{EOP}=\partial_x \rho_\textrm{f} \partial_y \phi\!-\!\partial_y \rho_\textrm{f} \partial_x \phi
\end{equation}
is identical to zero. This remains valid for any ion distribution for which $\rho_\textrm{f} = f(\phi)$ and with $\phi$ denoting the only variable. For instance, this also includes all systems that can be described by the Debye-H\"uckel (DH) approximation in the limit of low $\phi$. Hence, within the PB theory, for stationary and isothermal electrokinetic systems with vanishing inertial effects and no external field, the stream function is determined by the biharmonic equation $\bnabla^4 \psi_\textrm{PB} = 0$ \cite{Levich:Prentice1962,Pascall:JFM2011,Dietzel:JFM2017}. In general, the charge density cannot be expressed by $\rho_\textrm{f,PB}$ so that the EOP may be non-vanishing even without the application of an external electric field, indicating that the ion cloud in the vicinity of the wall is not in a state of mechanical equilibrium.

Expression (\ref{Eq:NSE_stream}) is made dimensionless by scaling the $x$- and $y$-directions by the SAW wavelength $\lambda_\textrm{SAW}$ and the channel height $h$, respectively, i.e., $\overline{x} = x/\lambda_\textrm{SAW}$ and $\overline{y} = y/h$. The aspect ratio is given by $A = h/\lambda_\textrm{SAW} = h f_\textrm{SAW}/c_\textrm{SAW}$. Furthermore, employing a scaling velocity denoted by $u_0$, the dimensionless stream function is defined by $\overline{\psi} = \psi/(h u_0)$, while $\overline{\omega} = \omega h/u_0$. Time is scaled using $f_\textrm{SAW}$ according to $\overline{t} = t f_\textrm{SAW}$. This leads to
\begin{eqnarray} 
\label{Eq:NSE_stream_nondim}  Ro \partial_{\overline{t}} \overline{\omega} + Re (\partial_{\overline{y}} \overline{\psi} \partial_{\overline{x}} \overline{\omega} - \partial_{\overline{x}} \overline{\psi} \partial_{\overline{y}} \overline{\omega}) = \\ \nonumber
(A^2 \partial^2_{\overline{x}} + \partial^2_{\overline{y}}) \overline{\omega} + Ha (\partial_{\overline{x}} \overline{\rho}_\textrm{f} \partial_{\overline{y}} \overline{\phi} - \partial_{\overline{y}} \overline{\rho}_\textrm{f} \partial_{\overline{x}} \overline{\phi}).
\end{eqnarray}
The dimensionless charge density is given by $\overline{\rho}_\textrm{f} = \rho_\textrm{f}/(e \nu n_0)$, while $\overline{\phi} = \phi e\nu/(k_\textrm{B} T)$. The (viscous) Roshko number, the Reynolds number, and the Hartmann number are defined by $Ro = h^2 f_\textrm{SAW} \rho/\eta$, $Re = A \rho u_0 h/\eta$, and $Ha = A h n_0 k_\textrm{B} T/(\eta u_0)$, respectively.

The non-dimensional Poisson equation is given by
\begin{equation} 
\label{Eq:Poisson_nondim} (A^2\partial^2_{\overline{x}} + \partial^2_{\overline{y}}) \overline{\phi} = -\frac{\overline{\kappa}^2_0}{2}(\overline{n}_+ - \overline{n}_-),
\end{equation}
where the non-dimensional Debye parameter is defined by $\overline{\kappa}_0 = h/\lambda_\textrm{D}$. The non-dimensional ion number concentrations $\overline{n}_\pm = n_\pm/n_0$ are governed by the non-dimensional NPEs, given by
\begin{align} 
\label{Eq:NPE_nondim} Ro_{\textrm{i},\pm} \partial_t \overline{n}_\pm +& Pe_{\textrm{i},\pm}(\partial_{\overline{y}} \overline{\psi} \partial_{\overline{x}} \overline{n}_\pm - \partial_{\overline{x}} \overline{\psi} \partial_{\overline{y}} \overline{n}_\pm) \\ \nonumber
=&(A^2 \partial^2_{\overline{x}} + \partial^2_{\overline{y}}) \overline{n}_\pm \pm \nu \overline{n}_\pm (A^2 \partial^2_{\overline{x}} + \partial^2_{\overline{y}}) \overline{\phi} \\ \nonumber
\pm& \nu (A^2 \partial_{\overline{x}} \overline{n}_\pm \partial_{\overline{x}}\overline{\phi} + \partial_{\overline{y}} \overline{n}_\pm \partial_{\overline{y}} \overline{\phi}),
\end{align}
where the ionic Roshko number and the ionic P\'eclet number are given by $Ro_{\textrm{i},\pm} = h^2 f_\textrm{SAW}/D$ and $Pe_{\textrm{i},\pm} = A u_0 h/D$, respectively.

This work focuses on flows driven by a non-vanishing EOP alone. In such systems, a non-uniform osmotic pressure is the main source for fluid motion and balanced by viscous stresses of the same order of magnitude. Hence, according to its definition, $Ha=1$ is assumed and used to define the scaling velocity according to $u_0 = A h n_0 k_\textrm{B} T/\eta = h^2 n_0 k_\textrm{B} T f_\textrm{SAW}/(c_\textrm{SAW} \eta)$. In this study, the thermophysical properties of an aqueous electrolyte solution and the operation parameters as summarized in Table \ref{Tbl:thermoprops_operation} are used.
\begin{table}
	\begin{center}			
		\begin{tabular}{l|c||l|c}
			\multicolumn{2}{c||}{Fluid properties} & \multicolumn{2}{c}{Other parameters} \\ \hline
			Parameter & Range of variation & Parameter & Range of variation \\ \hline
				&  &  & \\
			$\rho_0\:(\textrm{kg m}^{-3})$ & $1000$ & $c_\textrm{SAW}\:(\textrm{m s}^{-1})$ & $3965$$^{\textrm{a}}$ \\
			$\eta_0\:(\textrm{Pa s})$ & $1 \times 10^{-3}$ & $f_\textrm{SAW}\:(\textrm{MHz})$ & $10^{-1}-10^{2}$ \\
			$n_0\:(\textrm{M})$ & $(0.1-10) \times 10^{-3}$ & $\lambda_\textrm{SAW}\:(\textrm{m})$ & $10^{-5}-10^{-2} $ \\
			$D\:(\textrm{m}^2\:\textrm{s}^{-1})$ & $(1-5)\times 10^{-9}$ & $\hat{U}\:(\textrm{V})$ & $0.25-1.25$ \\
			$\epsilon/\varepsilon_0$ & $78.14$ $^{\textrm{b}}$ & $h\:(\textrm{m})$ & $(3-480) \times 10^{-9}$ \\
			$\nu$ & $1$ & $\lambda_\textrm{D}\:(\textrm{m})$ & $(3-30) \times 10^{-9}$ \\
		\end{tabular}
	\end{center}
\caption{Thermophysical properties of aqueous electrolyte solution and other parameters employed in this study. a - \citet{Yeo:Biomicrofluidics2009}, b - \citet{Buchner:PhysChemA1999}.}
\label{Tbl:thermoprops_operation}
\end{table}
With these values, the characteristic velocity and numbers pertinent to the present study can be computed as listed in Table \ref{Tbl:scaling}.
\begin{table}
	\begin{center}			
		\begin{tabular}{l|c||l|c}
			\multicolumn{4}{c}{Scaling parameters} \\ \hline
			Parameter & Range of variation & Parameter & Range of variation \\ \hline
				&  &  & \\
			$u_0\:(\textrm{m s}^{-1})$ & $10^{-11}-10^{-1}$ & $Re$ & $10^{-20}-10^{-4}$ \\
			$A$ & $10^{-8}-10^{-2}$ & $Ro$ & $10^{-5}-10^{1}$ \\
			$\overline{\kappa}_0$ & $10^{0}-10^{2}$ & $Pe_{i,\pm}$ & $10^{-18}-10^{-1}$ \\
			$Ha$ & $1$ & $Ro_{i,\pm}$ & $10^{-1}-10^{3}$ \\
		\end{tabular}
	\end{center}
\caption{Typcial values of the characteristic velocity and dimensionless numbers pertinent to the present study.}
\label{Tbl:scaling}
\end{table}
Hence, flow velocities corresponding to $Ha=1$ are far below $1\:\textrm{m s}^{-1}$. For narrow channels as considered in this work, $Re \ll 1$, i.e., advective momentum transport is negligibly small, while the unsteady term in Eq. (\ref{Eq:NSE_stream_nondim}) proportional to $Ro$ is not. With respect to the NPEs, in this work $Pe_{\textrm{i},\pm} = {\cal O}(10^{-7}-10^0)$, while $Ro_{\textrm{i},\pm}$ is typically several orders of magnitude larger. Hence, the ion transport governed by the NPEs is decoupled from the momentum transport. As a side note, for systems with an ion cloud pushed away from mechanical equilibrium by a strong externally driven flow characterized by a large $u_0$, counterintuitively, the corresponding EOP is independent of $u_0$. As can be seen from Eq. (\ref{Eq:NSE_stream_nondim}), the EOP is proportional to $Ha$, while $\overline{\rho}_\textrm{f}$ is governed by Eq. (\ref{Eq:NPE_nondim}) and proportional to $Pe_{\textrm{i},\pm}$. Hence, the EOP scales as $Ha Pe_{\textrm{i},\pm} = h n_0 k_\textrm{B} T/(\eta D)$, which is independent of $u_0$. Furthermore, since the base flow is proportional to $Re$ and scales as the EOP linearly with respect to $h$, enlarging $u_0$ or $h$ will not enhance the magnitude of the secondary flow driven by the EOP in comparison to the base flow velocity that causes the non-equilibrium EDL in the first place. Thus, it can be concluded that for an externally imposed base flow the secondary EOP-driven flow can practically always be disregarded, except for the special case of unusually small ion diffusivities.

Standing (SD) or traveling (TV) SAWs are considered in this work. The corresponding electrostatic wall potentials of amplitude $\hat{U}$ are assumed to be given by
\begin{equation} 
\label{Eq:Wallpot_SW} U_\textrm{SD} = \hat{U} \textrm{sin}(2 \pi \overline{t})\textrm{cos}(2\pi \overline{x} + \Delta \varphi)
\end{equation}
and
\begin{equation} 
\label{Eq:Wallpot_TW} U_\textrm{TV} = \hat{U} \textrm{sin}[2 \pi(\overline{x} - \overline{t}) + \Delta \varphi],
\end{equation}
respectively. The phase shift between two SAWs is denoted by $\Delta \varphi$, which is relevant for the cases where SAWs of identical frequency are applied on both channel walls. As will be discussed later on, the application of SAWs of different frequencies were found to be not beneficial to maximize the induced EOF transport.

The wall potential is screened not only by the diffuse part of the EDL but also by the immobile ions in the Stern layer. To account for this effect, the Stern layer model described by \citet{Olesen:PRE2010} is commonly implemented. It considers the Stern layer as a parallel-plate capacitor of capacitance $C_\textrm{St}$, while the continuity of the dielectric displacement field vector at the interface between the Stern and the diffusive layer leads to the following mixed boundary condition \cite{Olesen:PRE2010}
\begin{equation} 
\label{Eq:Stern} C_\textrm{St}(U-\phi)+\epsilon \boldsymbol{n}\cdot\bnabla \phi=0.
\end{equation}
The surface normal pointing into the electrolyte is denoted by $\boldsymbol{n}$. In non-dimensional form and with $\overline{U} = U e \nu/(k_\textrm{B} T)$, Eq. (\ref{Eq:Stern}) reads
\begin{equation} 
\label{Eq:Stern_nondim} \overline{U}-\overline{\phi} + \frac{\delta_\textrm{St}}{\overline{\kappa}} \overline{\boldsymbol{n}} \cdot \bnabla \overline{\phi}=0.
\end{equation}
For the parallel-plate channel $\overline{\boldsymbol{n}} \cdot \bnabla \overline{\phi} = \mp \partial_{\overline{y}} \overline{\phi}$, with the minus sign valid for the lower wall and the plus sign for the upper wall. The ratio between the (nominal) capacitance of the diffuse part of the EDL relative to the one of the Stern layer is denoted by $\delta_\textrm{St} = C_\textrm{D}/C_\textrm{St}$ with $C_\textrm{D} = \epsilon/\lambda_\textrm{D}$. While the model of the Stern layer has been incorporated in the mathematical framework of this work, the results discussed in the following have been obtained without it. Since it is a simple capacitor model, it just lowers the effective $\zeta$ potential acting on the fluid. This was verified by several numerical tests (not shown). Commonly used as a parameter to fit model results to those obtained from experiments, values of $\delta_\textrm{St}$ are difficult to estimate and may be in the range of $\delta_\textrm{St} = 0.01 - 10$ \cite{Olesen:PRE2010}. Furthermore, recent work suggests that due to extraordinarily slow ad- and desorption processes it may take a few hundred seconds to fully charge the Stern layer \cite{Werkhoven:PRL2018}. This would be orders of magnitude too long to follow the electric signal of the SAW. Thus, it is likely that for the high frequencies considered in this work the Stern layer remains practically uncharged. To avoid building our results on a boundary condition that is speculative in the present context, we neglect the Stern layer in most of our simulations. Instead, to obtain results that are conservative with respect to the flow velocities achieved, we assume a voltage amplitude not exceeding $1.25\:\textrm{V}$, which is approximately $30-40$ times smaller than typical values used for IDTs in practice \cite{Lei:LabChip2014}.

The voltage amplitude is attenuated with increasing distance from the IDT. If the voltage attenuation is proportional to the attenuation of the acoustic wave, one may write \cite{Dentry:PRE2014}
\begin{equation} 
\label{Eq:attenuation_volt} \hat{\overline{U}} = \hat{\overline{U}}_{|\overline{x}=0} \textrm{e}^{-\overline{\alpha}\; \overline{x}},
\end{equation}
with the (non-dimensional) attenuation coefficient given by
\begin{equation} 
\label{Eq:attenuation_coeff} \overline{\alpha} = \frac{\rho c_\textrm{liq}}{\rho_\textrm{s} c_\textrm{SAW}},
\end{equation}
with $\rho_\textrm{s}$ denoting the density of the piezo-active channel walls. Equation (\ref{Eq:attenuation_coeff}) is an accurate estimate only for an acoustic wave propagating in a solid that is in contact with a semi-infinite liquid body. For narrow channels with their increased viscous dissipation, it may serve only as a rough orientation. Herein, $\overline{\alpha}^{\:-1} \approx 12$, i.e., the voltage signal attenuates to $1/e$ of its initial value on a length that is about $12$ times larger than $\lambda_\textrm{SAW}$. This suggests that for channels that measure only a few multiples of $\lambda_\textrm{SAW}$ in length, attenuation may be neglected.

Large wall potentials can induce large ion densities, eventually leading to ion crowding \cite{Bazant:AdvCollIntScie2009,Bazant:PRL2011}. With respect to the latter, steric effects can be accounted for by employing the modified NPEs of the Bikerman model \cite{Kilic:PRE2007b}, which can be obtained by adding
\begin{align} 
\label{Eq:NPE_ioncrowd_nondim} \overline{j}^{\textrm{B}}_\pm = A^2 &\partial_{\overline{x}} \left[\overline{n}_\pm\frac{\partial_{\overline{x}}(\overline{n}_+ +\overline{n}_-)}{\overline{n}_\textrm{max} - \overline{n}_+ -\overline{n}_-}\right] \\ \nonumber
+ &\partial_{\overline{y}} \left[\overline{n}_\pm\frac{\partial_{\overline{y}}(\overline{n}_+ +\overline{n}_-)}{\overline{n}_\textrm{max} - \overline{n}_+ -\overline{n}_-}\right]
\end{align}
to the RHS of Eq. (\ref{Eq:NPE_nondim}). The maximal non-dimensional number concentration is given by $\overline{n}_\textrm{max} = 1/(a^3_\textrm{ion} n_0)$, where $a_\textrm{ion}$ denotes the distance between densely packed ions. Herein, $a_\textrm{ion} = {\cal O}(10^0-10^1)\:$\AA$\:$ \cite{Greberg:JChemPhys1998,Kilic:PRE2007a,Kilic:PRE2007b}, while $n_0 = {\cal O}(10^{-4}-10^{-2})\textrm{M}$, so that $\overline{n}_\textrm{max} = {\cal O}(10^3-10^6)$. Hence, in the light of the relatively low bulk ion concentrations assumed in this work, corrections obtained from Eq. (\ref{Eq:NPE_ioncrowd_nondim}) are negligibly small for most cases addressed in this study. This was verified numerically (not shown).

For sufficiently small potentials, the model outlined above can be linearized to provide an analytical expression for the time-averaged EOF velocity just outside of the EDL, which was derived by \citet{Gonzalez:PRE2000} for ACEOF. This (effective slip) velocity is given by
\begin{equation} 
\label{Eq:ACICEOF_vel} \langle u\rangle = -\frac{\epsilon}{4 \eta} \Lambda \partial_x |\phi-U|^2,
\end{equation}
where $\Lambda = 1/(1+\delta_\textrm{St})$. Using the linearized and time-averaged expressions for the charge density and the electric potential as derived in that work, one finds $|\langle \partial_y \rho_f \partial_x \phi \rangle| = |\langle \partial_x \rho_f \partial_y \phi \rangle| + \Lambda \partial_x |\phi-U|^2/4$. Thus, $F_\textrm{EOP}$ as expressed by Eq. (\ref{Eq:EOP}) is indeed non-vanishing for ACEOF. In a notation using primitive variables, this implies that $||\langle \bnabla p \rangle|| \neq || \langle \rho_f \bnabla \phi \rangle ||$. As illustration of this imbalance, one may assume that in the normal direction to the wall the pressure gradient and the gradient of the electrostatic stress exactly cancel each other since no flow through the wall is permitted. Hence, $\langle \partial_y p \rangle \approx -\langle \varepsilon \partial_y (\partial_y \phi)^2/2\rangle$ so that $p$ is fixed except for an integration constant. Since the pressure is a scalar, it acts with the same strength in the direction parallel to the channel wall. In that direction, there is no obstacle or boundary to enforce the balance between the pressure gradient and electrostatic stress so that in general $\langle -\partial_x p - \rho_f \partial_x \phi \rangle \neq 0$. This imbalance causes the net flow. In the lower part of Fig. \ref{Fig:sketch_flowdom}, this basic mechanism is illustrated for a wall subjected to a sinusoidal wave of the electrostatic surface potential. Despite the fact that the electric potential $\phi$ (indicated by a color scale) at the non-dimensional time $\overline{t} = 0.75$ is reversed in comparison to the instant at $\overline{t} = 0.25$, the corresponding fluid stress $-\bnabla p - \rho_f \bnabla \phi$ (indicated by white arrows) is identical and induces a fluid motion in the same direction (indicated by black arrows and black streamlines). Thus, despite the alternating polarity of the electrostatic potential one obtains (on time-average) an uni-directional flow. Based on Eq. (\ref{Eq:ACICEOF_vel}), the maximal velocity of such a fluid motion occurs at a frequency of 
\begin{equation} 
\label{Eq:ACICEOF_freqmax} f_{\textrm{max}} = \Lambda^{-1} \frac{0.199\sqrt{\sigma}}{\Delta x},
\end{equation}
with $\sigma = \epsilon D/\lambda^2_\textrm{D}$ denoting the electric conductivity of the electrolyte, while $\Delta x$ denotes the size of the gap between two electrodes. As it was suggested to be a valid assumption by experiments preceding the theoretical work \cite{Green:PRE2000}, in the course of the derivation of Eqns. (\ref{Eq:ACICEOF_vel}) and (\ref{Eq:ACICEOF_freqmax}) it was assumed that $f_{\textrm{max}} = {\cal O}(t^{-1}_\textrm{RC})$ \cite{Gonzalez:PRE2000}.

Herein, we do not consider stationary electrodes, but the surface potential is induced by SAWs standing or traveling along the wall. Equivalent systems have been considered in the context of ACEOF by placing several electrodes with a spacing of $\lambda_0 = 2 \pi/k_0$ along the wall and energizing them according to a traveling wave, where $k_0$ denotes the wave number of the wall electrode. In that case, the (time-averaged) electro-osmotic slip velocity just outside of the EDL is found to read \cite{Ramos:JAP2005}
\begin{equation} 
\label{Eq:TWACEOF_velmax} \langle u\rangle = \Lambda \frac{\epsilon k_0 \hat{U}^2}{2 \eta} \frac{\Omega_0}{1+\Omega^2_0},
\end{equation}
which is maximal at $\Omega_0 = 1$ with $\Omega_0 = 2 \pi f_0 C_\textrm{D}/(\sigma k_0)$. For SAW actuation, $k_0$ is not a free parameter but given by the dispersion relation of the SAW. Hence, by contrast to TV-ACEOF, one finds two opposing effects: Higher frequencies imply reduced charging of the EDL but larger wave numbers, while smaller frequencies imply stronger EDL charging but smaller wave numbers. More specifically, invoking the SAW dispersion relation to estimate the magnitude of the SAW-induced EOF implies that $\Omega_0 = c_\textrm{SAW} C_\textrm{D}/\sigma$ is a constant and that $\langle u\rangle$ as expressed by Eq. (\ref{Eq:TWACEOF_velmax}) is linearly increasing with the SAW frequency. Hence, no finite frequency with maximal EOF should be expected. With these considerations and for $c_\textrm{SAW} =  3965\:\textrm{m s}^{-1}$, $\hat{U} = 1\:\textrm{V}$ as well as the thermophysical properties listed in Table \ref{Tbl:thermoprops_operation}, one finds sizable SAW-EOF velocities above $1\:\mu\textrm{m s}^{-1}$ only for frequencies above $100\:\textrm{MHz}$. Ignoring the functional dependence of (\ref{Eq:TWACEOF_velmax}) on $\Omega_0$ and assuming  $f_\textrm{SAW} = 1\:\textrm{MHz}$ leads to an estimate of $\langle u\rangle_\textrm{SAW,max} = {\cal O}(10^{-1})\:\textrm{mm s}^{-1}$. Such an estimate corresponds to the Helmholtz-Smoluchowski (HS) velocity expressed by
\begin{equation} 
\label{Eq:Helmholtz_Smoluchowski} u_\textrm{HS} = \frac{\epsilon \zeta}{\eta} E,
\end{equation}
for which the electric field $E$ equals $2 \zeta/\lambda_\textrm{SAW}$, while $\zeta \equiv \hat{U}$. On the one hand, since ACEOF and SAW-induced EOF share the same physical origin, expression (\ref{Eq:TWACEOF_velmax}) is physically justified but provides unrealistic estimates. On the other hand, expression (\ref{Eq:Helmholtz_Smoluchowski}) provides more realistic estimates but the simplification leading to it cannot be fully justified. These contradicting estimates obtained from the linearized model ask for a more careful view on the interplay between the RC time of the EDL, high electrostatic surface potentials, and the dispersion relation of the SAW. As described in the next section, this can be addressed by a full nonlinear numerical simulation.

\section{Details of the computational model and test cases} \label{Sec:CompModel}

A sketch of the (non-dimensional) computational domain is shown in the upper part of Fig. \ref{Fig:sketch_flowdom}. All of the three segments $1-3$ have a non-dimensional height of one, where segments $1$ and $3$ are passive segments of length three that are not subjected to a SAW. Along the side walls of these segments (marked by "$c$"), no-slip, 
\begin{equation} 
\label{Eq:No-Slip} \partial_{\overline{x}} \overline{\psi} = \partial_{\overline{y}} \overline{\psi} = 0,
\end{equation}
and no-flux,
\begin{equation} 
\label{Eq:No-Flux} \overline{\boldsymbol{j}}_\pm \cdot \boldsymbol{n} = 0,
\end{equation}
are applied as boundary conditions, with $\boldsymbol{n}$ denoting the surface normal and
\begin{equation} 
\label{Eq:Non-Dim-Flux} -\overline{\boldsymbol{j}}_\pm = \overline{\bnabla} \overline{n}_\pm \mp \nu \overline{n}_\pm \overline{\bnabla}\:\overline{\phi}
\end{equation}
denoting the diffusive-electromigrative ion flux vector. Herein, the non-dimensional nabla operator is defined by $\overline{\bnabla} = (A \partial_{\overline{x}},\partial_{\overline{y}})^\textrm{T}$. At the channel ends (marked by "$d$"), while still enforcing the no-slip condition, the no-flux condition is replaced by the condition of periodicity of the ion concentrations between both channels ends. However, it was verified that enforcing the no-flux condition instead leads to the same results. Considering that the investigation of SAW-induced flow through a channel is the focus of the present work, it might be surprising that the no-slip boundary condition is used at the channel ends. As will be detailed further below, this is a consequence of the numerical solution approach. Segment $2$ incorporates no-slip and no-flux but piezo-active side walls (marked by "$a$" and "$b$"), where the electric boundary conditions expressed by Eqns. (\ref{Eq:Wallpot_SW}) or (\ref{Eq:Wallpot_TW}) are imposed. While $\lambda_\textrm{SAW}$ itself varies with $f_\textrm{SAW}$ according to the dispersion relation, for every simulation and independent of the frequency the non-dimensional length of segment $2$ equals five, i.e., five SAW wavelengths are included within the computational domain. For all cases, $\overline{\psi}=0$ was set along boundary $a$. 

The governing, time-dependent equations (\ref{Eq:NSE_stream_nondim})-(\ref{Eq:NPE_nondim}) along with its boundary conditions were implemented using the partial differential equations (PDE) mode in COMSOL MULTIPHYSICS 5.4 \cite{Comsol2019} and solved on a dense structured but anisotropic grid, which was highly refined along the charged walls to improve the resolution of the local electric field. Quadratic (Lagrangian) shape functions were employed for the spatial discretization, while the constant Newton-Raphson method was used to linearize the non-linear parts of the governing equations. All simulations were carried out on a Dell Precision T7500 workstation with Ubuntu 12.04 LTS as operating system. Appropriate convergence studies were carried out with respect to mesh density and time step size. Since the COMSOL solver employs a variable time-stepping scheme according to a specified relative tolerance, the temporal convergence study was conducted by successively tightening the relative residual during each time step iteration, which forces the iterative solver to reduce the time integration increments.
\begin{figure}
	\centerline{\includegraphics[width=12cm]{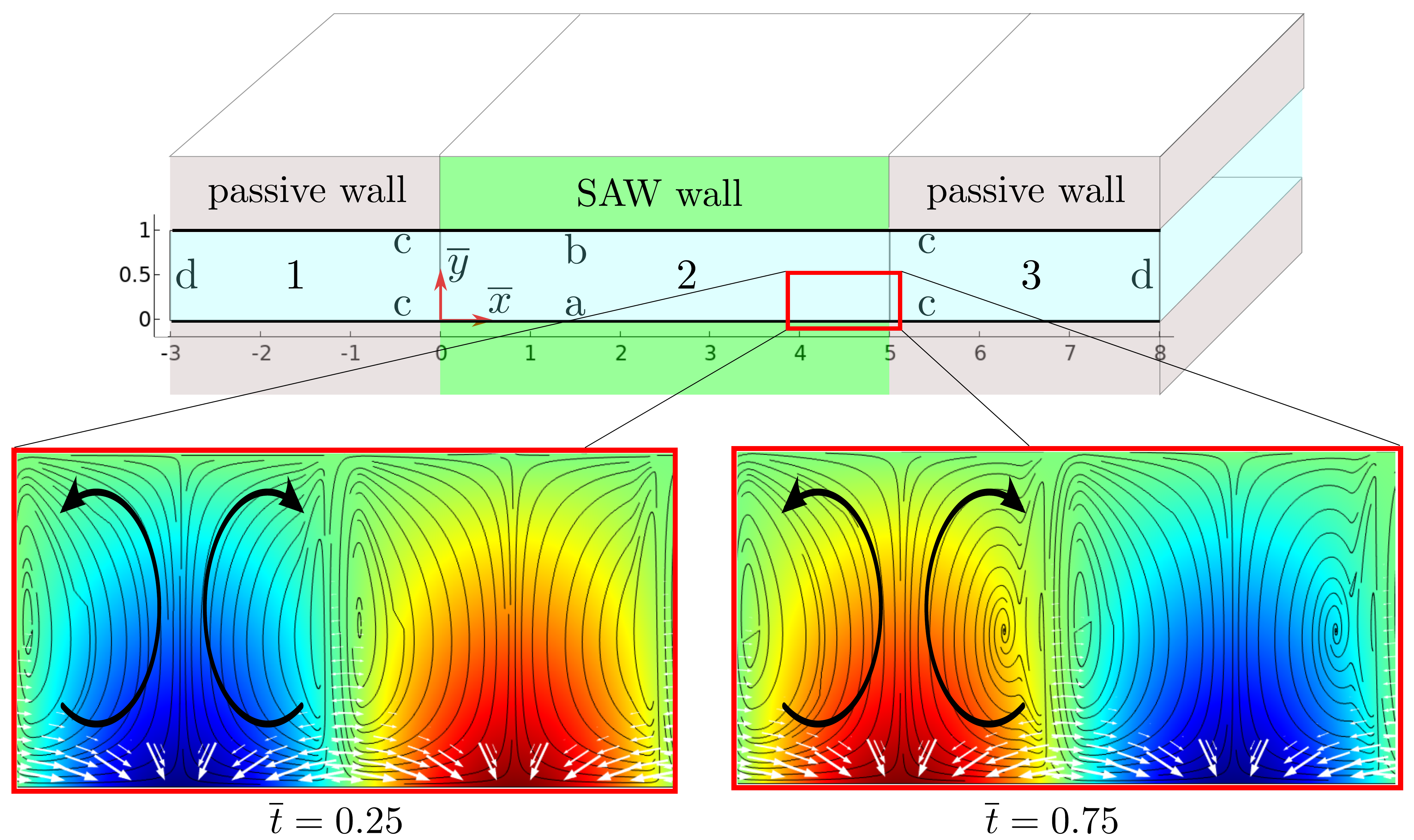}}
	\caption{In the upper part of this figure, a sketch of the parallel-plate channel, the computational domain, and the employed boundary conditions are shown. Segments $1$ and $3$ are passive segments not subjected to a SAW, employing no-slip boundaries of unspecified electric potential, supplemented by either a no-flux boundary condition (boundaries c) or periodic ion concentrations (boundaries d). Segment $2$ consists of no-slip and no-flux but piezo-active side walls (boundaries a and b). In the lower part, the electric potential $\phi$ (color scale online, gray scale in print), the (polarity-independent) fluid stress $-\bnabla p - \rho_\textrm{f} \bnabla \phi$ (white arrows), as well as the corresponding fluid motion (black arrows and black streamlines) are shown for the two (non-dimensional) time instants $\overline{t} = 0.25$ and $\overline{t} = 0.75$.}
	\label{Fig:sketch_flowdom}
\end{figure}

With the described stream function-vorticity formulation, it was not possible to simulate SAW-driven EOF in channels with open ends, for which the no-slip condition at the channel ends (segments "$d$") needs to be replaced by the condition of periodicity with respect to the (axial) flow velocities. In such cases, it was found that an artificial pressure-driven flow is superimposed, which strongly exceeds the EOF. The reason for this is that the advective term in Eq. (\ref{Eq:NSE_stream_nondim}) is typically vanishingly small so that the effective momentum equation is linear. In this case, an artificial pressure-driven flow described by $\overline{\psi}_\textrm{art}$ with $\bnabla^4 \overline{\psi}_\textrm{art}=0$ can be added to the EOF described by $\overline{\psi}=\overline{\psi}_\textrm{EOF}$, so that $\overline{\psi}=\overline{\psi}_\textrm{EOF}+\overline{\psi}_\textrm{art}$ is still a solution of the governing momentum equation. This problem could only be circumvented by simulating channels with closed ends, for which the SAW-driven EOF leads to a (now physically justified) pressure gradient along the channel and a corresponding (time-averaged) backflow with a Hagen-Poiseuille (HP) profile. For a parallel-plate channel as depicted in Fig. \ref{Fig:sketch_flowdom} and in a non-dimensional form, this is described by 
\begin{equation} 
\label{Eq:Hagen-Poiseuille_nondim} \overline{\psi}_\textrm{HP} = \frac{1}{2} A \partial_{\overline{x}} \overline{p} \left[\frac{1}{3}(\overline{y}^3-1)-\frac{1}{2}(\overline{y}^2-1) \right],
\end{equation}
where the non-dimensional pressure gradient along the channel is given by $A \partial_{\overline{x}} \overline{p}_\textrm{HP} = \partial_{\overline{y}} \overline{\omega}_{|\overline{y}=0.5}$. The SAW-driven EOF was computed by subsequently subtracting the analytical solution (\ref{Eq:Hagen-Poiseuille_nondim}) from the numerically obtained complete solution, where $\partial_{\overline{y}} \overline{\omega}_{|\overline{y}=0.5}$ was taken as the line-average along the channel center-plane of segment $2$ from the time-averaged numerical simulation. Note that for a pure EOF, this value is identical to zero. Attempts to prevent the artificial pressure-driven flow by imposing $\partial_{\overline{y}} \overline{\omega}_{|\overline{y}=0.5}=0$ along the center plane or along the domain boundaries were not successful.

To allow for a parametric study of $f_\textrm{SAW}$, it is essential for any modeling framework of SAW-induced flow that the domain length can be scaled by a multiple of $\lambda_\textrm{SAW}$ since $f_\textrm{SAW}$ affects $\lambda_\textrm{SAW}$ via the dispersion relation. Otherwise every other value of $f_\textrm{SAW}$ would require a different computational domain if always the same number of multiplies of $\lambda_\textrm{SAW}$ should fit into the domain. In the present study, employing a commercial fluid solver, this restricted the choice of the problem formulation and numerical implementation to the chosen one, at the expense of involving the problem of an artificial pressure-driven flow.

To check the numerical implementation, the simulation code was used to reproduce well-known solutions to relevant EOF problems. First, in the thin EDL limit and for constant, low-valued wall $\zeta$ potentials, it was ensured that at steady-state one obtains the HS velocity (\ref {Eq:Helmholtz_Smoluchowski}) and the corresponding (quasi) plug flow. Second, to check the correct implementation of the time-dependent terms, the numerical work of \citet{Suh:IJNumMethFluids2011} and \citet{Suh:CollSurfA2011} was used. In these studies, the transient flow induced by a sudden increase of the $\zeta$ potential from zero to a constant value is addressed. Starting from uniform ion concentrations, the flow vanishes as soon as the ion concentrations in wall vicinity follow the Boltzmann distribution such that the ion cloud in the EDL is in a state of mechanical equilibrium. Comparing the domain-averaged absolute flow velocity as a function of time for a wall potential of either $0.2\:\textrm{V}$ or $1\:\textrm{V}$ (as shown in Fig. 2 of Ref. \citet{Suh:CollSurfA2011}), full quantitative agreement was found.

The comparison with early numerical studies on ACEOF is hampered by the circumstance that most of them address systems with larger geometric length scales to allow for an easier experimental validation. For instance, in \citet{Green:PRE2002}, ACEOF is induced by two wall electrodes placed next to each other and actuated by AC voltages in the $\textrm{kHz}$ range, while the transport phenomena due to the non-equilibrium EDL are not resolved. Instead, a thin-double-layer model is used which imposes Eq. (\ref{Eq:ACICEOF_vel}) as an analytical expression for the time-averaged EOF velocity at the boundary between the diffuse layer and the bulk. The effective Debye parameters characteristic for such studies are of the order of $\overline{\kappa}_0 = {\cal O}(5 \times 10^3)$. The simulation code underlying our work fully resolves the non-equilibrium processes in the EDL both in space and time. To be applicable for larger systems, a high mesh density with strong grid anisotropy needs to be employed, next to high orders (at least cubic) of the interpolation functions used for discretization. Since \citet{Green:PRE2002} solely plot the simulated time-averaged streamlines without providing quantitative flow characterization, only a qualitative comparison was possible, which was conducted at a frequency of $1\:\textrm{kHz}$ (Fig. 11 (c) in Ref. \cite{Green:PRE2002}) and suggested good agreement. For quantitative tests, a comparison with the system simulated by \citet{Pribyl:InTech2008} was undertaken. In that work, ACEOF is induced by AC voltages with frequencies ranging from $10^1$ to $10^5\textrm{Hz}$, while the fully resolved EDL is characterized by an effective Debye parameter of $100$. For example, for a selected AC frequency of $100\:\textrm{Hz}$, the evolution of the cross-averaged axial velocity in time as shown in Fig. 2 of Ref. \cite{Pribyl:ISEHD2006} was fully reproduced with the present code.

All simulations discussed in the following start at $\overline{t}=0$ with a uniform bulk ion concentration of $\overline{n} = (\overline{n}_+ + \overline{n}_-)/2 = 1$. As such an initial bulk ion distribution at non-vanishing $\zeta$ potentials implies mechanical non-equilibrium, a transient osmotic flow is observed even if a time-independent wall potential is applied. This transient flow vanishes with time, and in the case of applying an AC voltage a stable periodic flow remains \cite{Pribyl:InTech2008}. To obtain reproducible time-averaged results, all simulations consisted of two parts: The first was undertaken until the flow field at $\overline{t}=\overline{t}_2$ was indistinguishable from the one at $\overline{t}=\overline{t}_1=\overline{t}_2 - 1$, i.e., one time period before. In this context, the existence of a flow field that is synchronized with the SAW indicates that subharmonic modes are not present, permitting averaging over one SAW time period. Consequently, a second simulation of the same case for a time period of $\Delta \overline{t}=1$ was conducted, with the results obtained from the previous one at $\overline{t}=\overline{t}_2$ as start values. Intermediate solutions within $\Delta \overline{t}$ were saved at narrow increments and subsequently used to compute the time-averaged flow field by numerical integration with high accuracy.

\section{Results and discussion} \label{Sec:param}

\subsection{Standing waves} \label{Sec:param_sd}

First, cases are considered in which the $\zeta$ potential along the walls of the channel mid segment $2$ (see Fig. \ref{Fig:sketch_flowdom}) follows a standing SAW expressed by Eq. (\ref{Eq:Wallpot_SW}) with the voltage amplitude amounting to $\hat{U} = 1.25\:\textrm{V}$. Either a single wave (SW) on one side or two waves (DW) of the same frequency but with a phase shift $\Delta \varphi$ on both sides of that channel segment are employed. In Fig. \ref{Fig:STDSAW_vel_rms}, the space- and time averaged root-mean-square (rms) velocity defined by
\begin{equation} 
\label{Eq:vel_rms} \langle v \rangle_\textrm{rms} = \frac{u_0}{5} \int^1_0 \int^5_0 \sqrt{\partial_{\overline{x}} \langle\overline{\psi} \rangle^2 + \partial_{\overline{y}} \langle \overline{\psi} \rangle^2} d \overline{x} d \overline{y},
\end{equation}
is shown as a function of the applied SAW frequency, with $\Delta \varphi$ as a parameter and where
\begin{equation} 
\label{Eq:time_avg} \langle\overline{\psi} \rangle = \int^1_0 \overline{\psi}d\overline{t}
\end{equation}
denotes the time average of the stream function over one period. While $\langle v \rangle_\textrm{rms}$ is non-vanishing, there is no net axial velocity along the channel, i.e., at every instant in time, the axial flow velocity integrated across the channel width is identical to zero. For a qualitative discussion, the streamlines at $\overline{t} = 0.25$ and $10\:\textrm{MHz}$ are shown on the RHS panel of Fig. \ref{Fig:STDSAW_vel_rms}, where the plotted domain height equals $h$, while its length equals $2 \lambda_\textrm{SAW}$. Qualitatively, the time-averaged streamlines for each case look identical to the instantaneous ones (not shown), except for a very brief moment in every quarter cycle when the instantaneous vortices change its sense of rotation. As a representative example, $\langle\overline{\psi} \rangle$ is shown for $f_\textrm{SAW} = 10\:\textrm{MHz}$ and $\Delta \varphi = \pi/2$ on the top panel of Fig. \ref{Fig:STDSAW_vel_rms}. From the stream function plots, it can be seen that the SAW-induced EOF may cause two fundamentally different flow patterns: For a SW or a DW with $\Delta \varphi = \pi/2$, the vortices stretch over the complete channel width, while for a DW with either $\Delta \varphi = 0$ or $\pi$ the vortices are symmetric with respect to the channel center plane. This implies that in these cases twice the number of vortices are present. In comparison to the SW case, for which the vortices are aligned vertically to the channel center plane, the vortices of the DW case with $\Delta \varphi = \pi/2$ are tilted, with the tilt angle being proportional to $A$. For the latter case, $\langle v \rangle_\textrm{rms}$ is up to three times larger than for the SW case. The values of $\langle v \rangle_\textrm{rms}$ of the DW case with $\Delta \varphi = 0$ are negligibly small. As implied by expression (\ref{Eq:Stern}), ICEOF is mainly driven by electric currents into the EDL. Since for $\Delta \varphi = 0$ the opposing channel walls are kept at a potential distribution being symmetric to the center plane, such cross currents are minimized and no measurable ICEOF is achieved. The largest value of $\langle v \rangle_\textrm{rms}$ is obtained at $10\;\textrm{MHz}$ and $\Delta \varphi=\pi/2$, amounting to $65\;\mu \textrm{m s}^{-1}$. By contrast to the expectation discussed before in the context of Eq. (\ref{Eq:TWACEOF_velmax}), there seems to exist an optimal actuation frequency, which lies between $1$ and $10\:\textrm{MHz}$ in all cases studied. In earlier work on ICEOF and ACEOF, larger systems were considered for which $t^{-1}_\textrm{RC}$ is in the kHz-range. Since narrow channels are considered herein, $t_\textrm{RC}$ is significantly smaller than in these previous cases. For $h = 480\:\textrm{nm}$, $\lambda_\textrm{D} = 9.61\:\textrm{nm}$, and $D = 5 \times 10^{-9}\:\textrm{m}^2\:\textrm{s}^{-1}$, one finds $t^{-1}_\textrm{RC} = 1.1\:\textrm{MHz}$. Hence, the maximum in the average velocity can be explained by the characteristic time scale to charge the electric double layer. Figure \ref{Fig:STDSAW_vel_rms} underpins not only the importance to utilize two SAWs on the opposing channel walls instead of just a single SAW, but also the sensitivity of the system with respect to $\Delta \varphi$: Given that no electric current enters or leaves the computational domain, the electric charge needed to rebuild the EDL on one side of the channel originates from the EDL on the opposite side. Consequently, the SAW-induced EOF goes along with large periodic currents connecting the EDLs on the opposite sides of the channel.
\begin{figure}
	\centerline{\includegraphics[width=12cm]{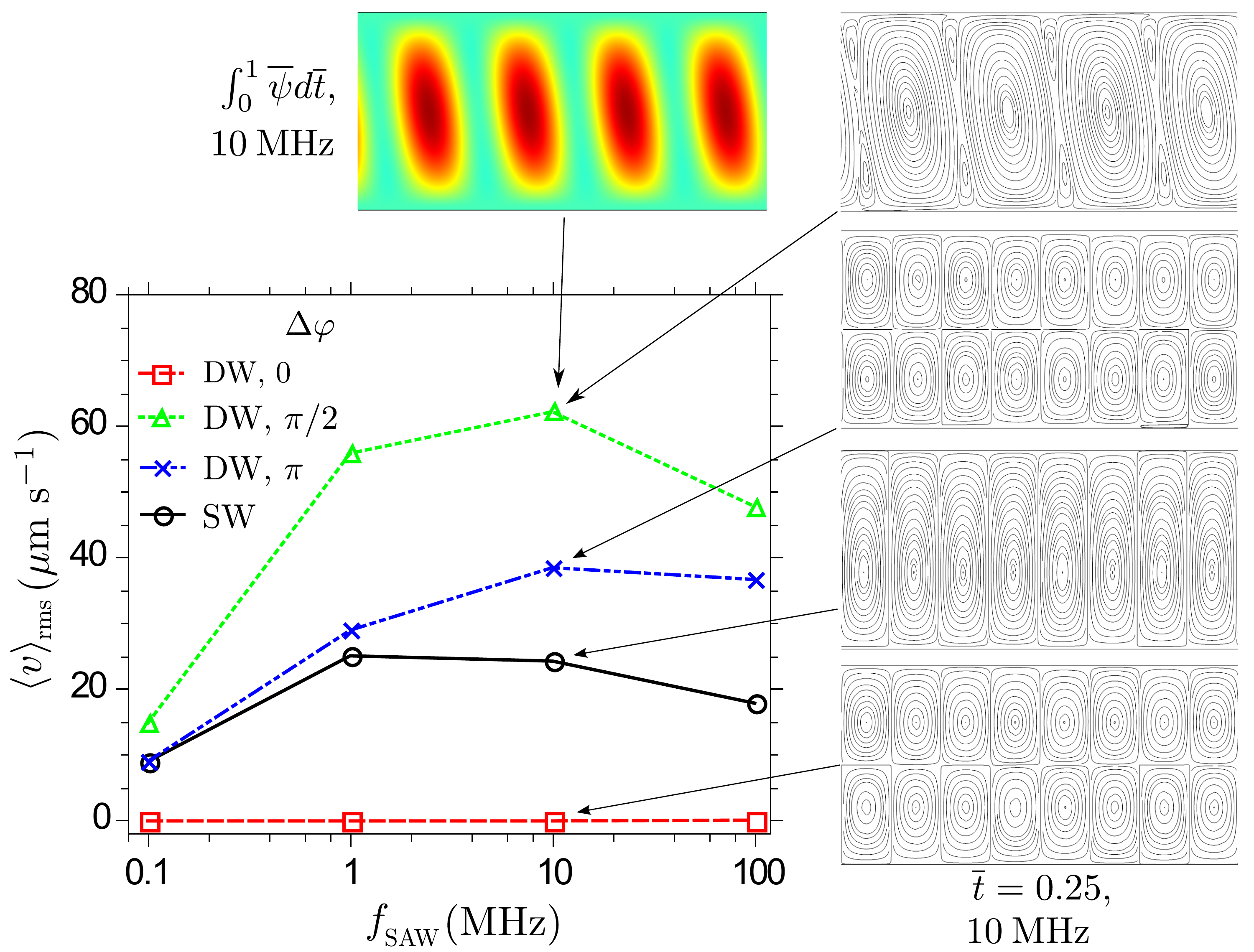}}
	\caption{Space- and time-averaged rms velocity $\langle v \rangle_\textrm{rms}$ within the channel mid segment $2$ defined in Fig. \ref{Fig:sketch_flowdom} are shown as a function of the SAW frequency. Either a single (SW) SAW is applied to the lower wall or two (DW) SAWs of the same frequency but with a phase shift $\Delta \varphi$ are applied on both sides of that channel segment. A $\zeta$ potential distribution was imposed as given by Eq. (\ref{Eq:Wallpot_SW}), where the voltage amplitude amounts to $\hat{U} = 1.25\:\textrm{V}$. Representative streamlines for the different SW and DW configurations are shown in the panels on the right hand side of the $x$-$y$ diagram. The time-averaged stream function $\int_0^1 \overline{\psi} d\overline{t}$ is exemplarily shown for $f_\textrm{SAW} = 10\:\textrm{MHz}$ and $\Delta \varphi = \pi/2$ on the top panel. The domain height, the nominal EDL thickness, and the bulk ion concentration equal $h = 480\;\textrm{nm}$, $\lambda_\textrm{D} = 9.61\:\textrm{nm}$, and $n_0 = 1\:\textrm{mM}$, respectively. The plotted domain length equals $2 \lambda_\textrm{SAW}$, where $\lambda_\textrm{SAW} = c_\textrm{SAW}/f_\textrm{SAW}$ with $c_\textrm{SAW} = 3965\:\textrm{m s}^{-1}$. For reference, $t^{-1}_\textrm{RC} = 1.1\:\textrm{MHz}$. The lines connecting the data points in the $x$-$y$ diagram are guides to the eye.}
	\label{Fig:STDSAW_vel_rms}
\end{figure}

For comparison, a number of simulations were conducted where the Stern layer was implemented via relation (\ref{Eq:Stern_nondim}) (not shown). The Stern parameter $\delta_\textrm{St}$ was varied from $0.1$ to $10$. It was found that with increasing $\delta_\textrm{St}$ the Stern layer reduces the effective $\zeta$ potential, weakening the SAW-induced EOF. For $\delta_\textrm{St} = 0.1$ and below, $\langle v \rangle_\textrm{rms}$ was practically unaffected by the presence of the Stern layer. Furthermore, given the large electric potentials involved, it was found that invoking the Debye-H\"uckel (DH) approximation to calculate the voltage drop across the Stern layer from (\ref{Eq:Stern}) by the simplified expression
\begin{equation} 
\label{Eq:Stern_volt_drop} \phi_{|y=0} \approx \frac{\hat{U}}{1+\delta_\textrm{St}}
\end{equation}
is inaccurate and of little practical relevance.

For the following discussion of traveling waves, it is emphasized that for a standing SAW the direction of rotation of the vortices depends on the sign of the temporal change of the local electric potential. For an increasing absolute local value, flow is directed towards such areas, while for a decreasing absolute local value, flow is directed away from it. This implies that in each cycle of the SAW, the vortices change their direction of rotation four times. During each vortex reversal, additional vortices emerge from areas close to the walls which subsequently replace the previous counter-rotating vortices. This process is illustrated in Fig. \ref{Fig:vortex_reversal}, where the height of the plotted domain equals $h$, while its length equals $\lambda_\textrm{SAW}$. From these plots, it is apparent that there are always four vortex pairs (each pair consisting of two counter-rotating vortices) within a single $\lambda_\textrm{SAW}$, which corresponds to the four sections of a sinusoidal wave in which the electric surface potential either increases or decreases.
\begin{figure}
	\centerline{\includegraphics[width=12cm]{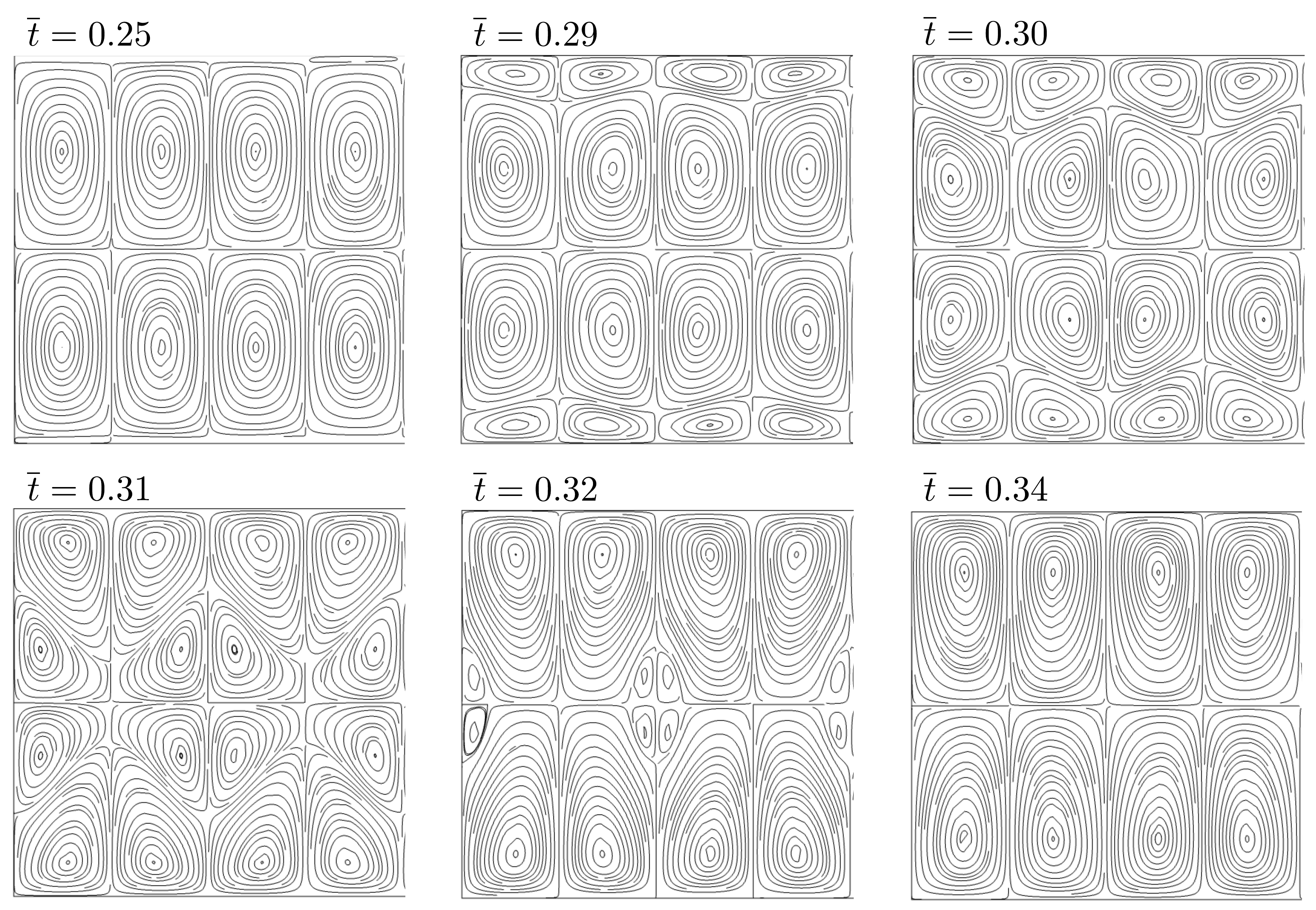}}
	\caption{Time evolution of the vortex pattern upon sign change of $d_t U_\textrm{SD}$ for the DW case with $f_\textrm{SAW} = 10\:\textrm{MHz}$, $\Delta \varphi=\pi$, and $\hat{U}=1.25\:\textrm{V}$. The nominal EDL thickness, the bulk ion concentration, as well as the plotted domain height and length equal $\lambda_\textrm{D} = 9.61\:\textrm{nm}$, $n_0 = 1\:\textrm{mM}$, $h = 480\:\textrm{nm}$, and $\lambda_\textrm{SAW} = 0.396\:\textrm{mm}$, respectively.}
	\label{Fig:vortex_reversal}
\end{figure}
A number of simulations were conducted where the SAW wavelength along the upper channel wall differs from the one at the bottom wall, potentially causing (frequency) beating effects (not shown). In all cases studied, the time-averaged velocities were always lower than for the corresponding case with identical wavelengths on both walls. Hence, if increasing the effective flow velocities is the key target, the SAW frequencies and wavelengths should be identical on both walls.

\subsection{Traveling waves} \label{Sec:param_tv}

The flow profiles induced by traveling SAW waves appear qualitatively identical to those shown in Fig. \ref{Fig:STDSAW_vel_rms} as insets, just that the vortex pairs are not stationary but travel along the channel. Hence, by contrast to a stationary SAW, a traveling SAW induces a net axial flow. Such flow profiles differ from those observed in TW-ACEOF, where the streamlines meander along the channel due to the presence of the electrodes but are not closed to form vortices \cite{Ramos:JAP2005,Yeh:PRE2011}. The reason for this is that in the latter case the electrode spacing is such that $A={\cal O}(1)$, while in the present case $A \ll 1$.

Figure \ref{Fig:TVLSAW_vel_rms} (a) shows the net axial velocity defined by
\begin{equation} 
\label{Eq:vel_ax} \langle u \rangle_\textrm{ax} = \frac{u_0}{5} \int^1_0 \int^5_0 |\partial_{\overline{y}} \langle\overline{\psi} \rangle - \partial_{\overline{y}} \overline{\psi}_\textrm{HP}| d \overline{x} d \overline{y}
\end{equation}
as a function of $f_\textrm{SAW}$ if two traveling waves move along both walls of channel segment $2$ with a phase shift of either $\Delta \varphi =\pi/2$ or $\pi$. As discussed before, due to the emergence of an artificial pressure-driven flow in simulations of a channel with open ends exposed to a traveling SAW, all simulations were conducted for a channel with closed ends. Since in that case the traveling SAW leads to a (physically justified) pressure-driven backflow, to arrive at the net axial flow velocity, in Eq. (\ref{Eq:vel_ax}) the corresponding HP flow profile is subtracted from the axial flow velocity computed numerically. The time-averaged stream function $\langle \overline{\psi} \rangle$ is calculated by the numerical simulations for a closed channel, while $\overline{\psi}_\textrm{HP}$ is given by Eq. (\ref{Eq:Hagen-Poiseuille_nondim}), along with the approach to obtain $A \partial_{\overline{x}} \overline{p}_\textrm{HP} = \partial_{\overline{y}} \overline{\omega}_{|\overline{y}=0.5}$ from the numerical simulations, as described before. For the simulations shown, $h=480\:\textrm{nm}$, while $\lambda_\textrm{D}=9.61\:\textrm{nm}$, and $\hat{U} = 1.25\:\textrm{V}$. The largest axial mean velocity is obtained for $f_\textrm{SAW} = 10\:\textrm{MHz}$ and $\Delta \varphi = \pi$, amounting to more than $0.2\:\textrm{mm s}^{-1}$. This is almost three times more than the largest value of $\langle v \rangle_\textrm{rms}$ obtained for standing waves under the same conditions. The reason for this discrepancy is the absence of vortex reversals in the TV wave cases. While for the SD wave cases the vortex pairs change their rotation direction four times in each cycle, the vortex pairs in the TV wave cases keep their direction of rotation within a reference frame co-moving with $\langle v \rangle_\textrm{ax}$.

In Fig. \ref{Fig:TVLSAW_vel_rms} (b), $\langle u \rangle_\textrm{ax}$ is shown as a function of $\overline{\kappa}_0 = h/\lambda_\textrm{D}$, with the nominal Debye length as a parameter. The latter is adjusted by employing different bulk ion concentrations in the range of $0.1\:\textrm{mM} \leq n_0 \leq 10\:\textrm{mM}$. For all simulations shown in this plot, $f_\textrm{SAW} = 10\:\textrm{MHz}$ and $\Delta \varphi = \pi$. In general, strong confinement as expressed by $\overline{\kappa}_0 \rightarrow 1$ leads to vanishing SAW-induced EOF, while those $\overline{\kappa}_0$ values where the peak velocity for a specific $n_0$ occurs are found to differ substantially from case to case and strongly depend on $\lambda_\textrm{D}$. Smaller $\lambda_\textrm{D}$ lead to larger $\langle u \rangle_\textrm{ax}$, reaching more than $1.3\:\textrm{mm s}^{-1}$ at $\lambda_\textrm{D} = 3.04\:\textrm{nm}$. This behavior can be explained by the enhanced gradients of the electric potential and charge density, causing an enhanced EOP. Conducting simulations with $\lambda_\textrm{D}$ reduced even further by increasing $n_0$ was not considered useful, as one would enter the regime of ion crowding. Since the Poisson-Nernst-Planck model considers the ions as point charges and is strictly valid only in the dilute limit, the validity of corresponding results would be questionable \cite{Bazant:AdvCollIntScie2009}. Nevertheless, using more accurate models incorporating such effects of ion crowding is an interesting path to be pursued further. Note in this context that any model describing the thin EDL limit with a framework based on the Poisson-Boltzmann theory is unsuitable, as the EOP vanishes in this case.
\begin{figure}
	\centerline{\includegraphics[width=12cm]{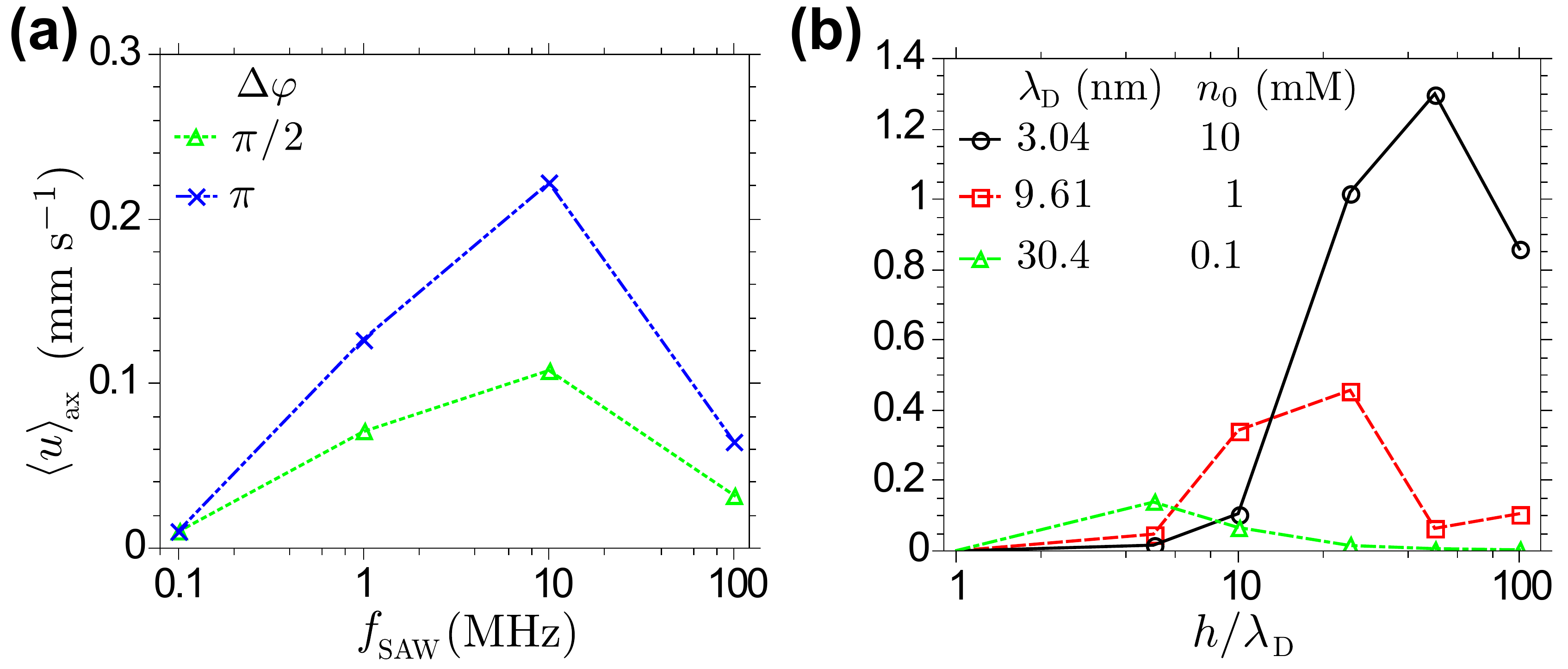}}
	\caption{EOF induced by traveling SAWs moving along the walls of channel segment $2$. In panel (a), the time- and space-averaged axial velocity $\langle u \rangle_\textrm{ax}$ as a function of $f_\textrm{SAW}$ is shown, with the phase shift $\Delta \varphi$ as a parameter, while $h=480\:\textrm{nm}$, $\lambda_\textrm{D}=9.61\:\textrm{nm}$, and $n_0 = 1\:\textrm{mM}$. For reference, $t^{-1}_\textrm{RC} = 1.1\:\textrm{MHz}$. In panel (b), $\langle u \rangle_\textrm{ax}$ as a function of  $h/\lambda_\textrm{D}$ is shown for $\Delta \varphi = \pi$, with the nominal Debye length $\lambda_\textrm{D}$ as a parameter, which corresponds to the bulk salinity noted next to it. For both plots $\hat{U} = 1.25\:\textrm{V}$. The lines connecting the data points are guides to the eye.}
\label{Fig:TVLSAW_vel_rms}
\end{figure}

Figure \ref{Fig:TVLSAW_uax} shows the time-averaged axial velocity profiles $\langle \overline{u}(\overline{y})\rangle = \langle u(y/h) \rangle/u_0$, corrected for the HP-backflow, with the frequencies and phase shifts as used in Fig. \ref{Fig:TVLSAW_vel_rms} (a) being the varied parameters. All other parameters are identical to those employed in the latter figure. In Fig. \ref{Fig:TVLSAW_uax} (a), $\Delta \varphi = \pi$ is used. All cases resemble the classical EOF plug-like flow profile, except for the transition from the flow inside the EDL and the bulk. Here, distinct peak velocities within the EDL can be observed, implying a lagging of the bulk flow that depends on the SAW frequencies. This effect becomes stronger with higher frequency so that for $100\:\textrm{MHz}$ the peak velocity exceeds the flow velocity at the channel center plane by up to $22\:\%$, while for frequencies lower than $100\:\textrm{kHz}$ it disappears. Hence, it is an inertial effect, where for higher frequencies, the flow cannot attain a quasi-steady state before the sign of the surface potential changes again. Figure \ref{Fig:TVLSAW_uax} (b) shows the results for $\Delta \varphi = \pi/2$. One observes that for increasing $f_\textrm{SAW}$ the axial net velocity profile becomes increasingly asymmetric with respect to the channel center plane. The instantaneous flow pattern for this case looks qualitatively similar to the corresponding SD case as shown in Fig. \ref{Fig:STDSAW_vel_rms}. The vortex pairs occupy not only the complete channel cross section but are also tilted; i.e., they are not mirror-symmetric with respect to the channel center plane. In addition, the clockwise and counter-clockwise rotating vortices differ in shape and magnitude. Time-averaging over such an instantaneous flow profile leads to the asymmetric net flow profile shown in Fig. \ref{Fig:TVLSAW_uax} (b).
\begin{figure}
	\centerline{\includegraphics[width=12cm]{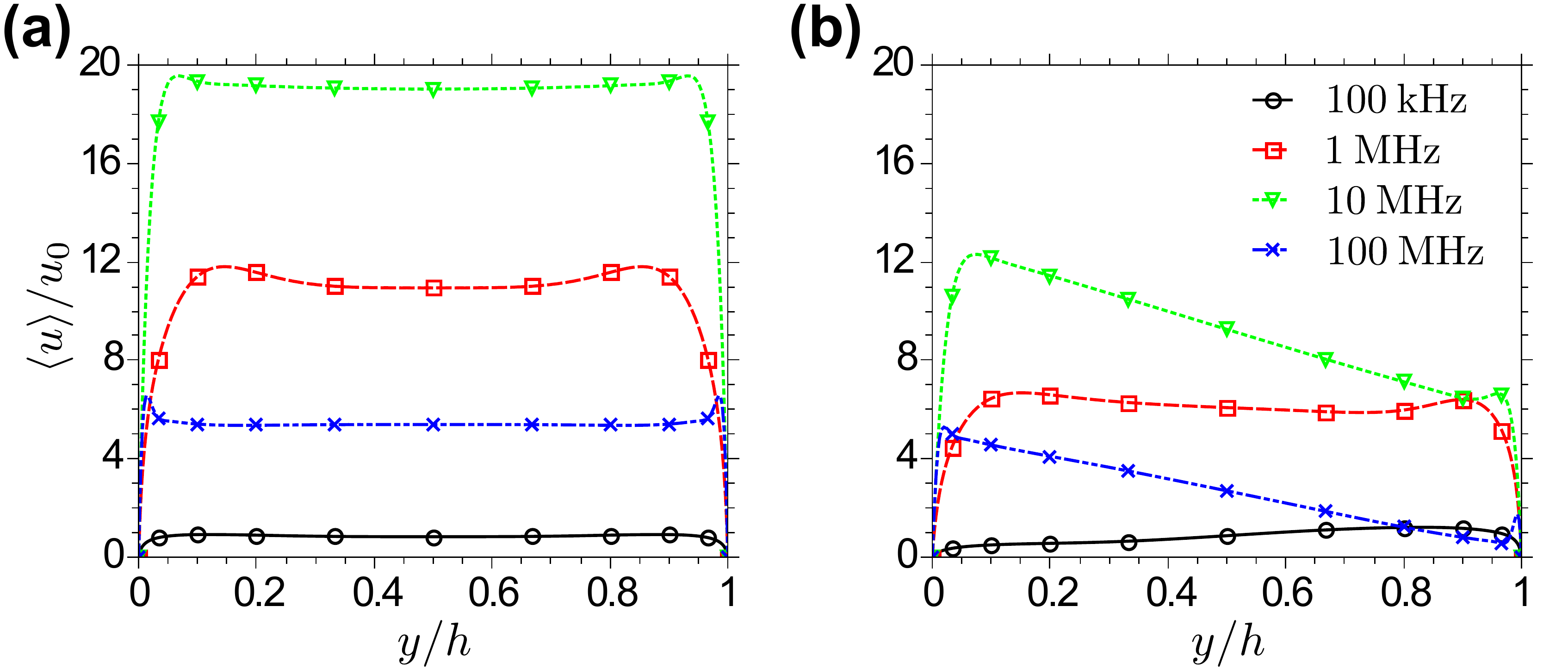}}
	\caption{Time-averaged axial velocity profiles $\langle \overline{u}(\overline{y})\rangle = \langle u (y/h) \rangle/u_0$ with $f_\textrm{SAW}$ as parameter. In panel (a) $\Delta \varphi = \pi$ while in panel (b) $\Delta \varphi = \pi/2$. In all simulations, $h = 480\:\textrm{nm}$, $\lambda_\textrm{D} = 9.61\:\textrm{nm}$, $n_0 = 1\:\textrm{mM}$, $\hat{U} = 1.25\:\textrm{V}$, and $u_0 = 11.9\:\mu\textrm{m s}^{-1}$. The lines connecting the data points are guides to the eye.}
\label{Fig:TVLSAW_uax}
\end{figure}

Some simulations were conducted (not shown) where the SAW on the top wall travels in the opposite direction to the one on the lower wall. For such a configuration, no axial net flow can be observed. Instead, a flow pattern emerges which resembles those found for two standing SAW with a phase shift of $\Delta \varphi = \pi/2$. While for the latter case the vortices change their sense of rotation four times in each cycle, the vortices caused by two counter-traveling SAWs span the complete channel cross section and keep their rotation direction. Consequently, large $\langle v \rangle_\textrm{rms}$ values of the order of $0.2\:\textrm{mm s}^{-1}$ ($10\:\textrm{MHz}$, $h=480\:\textrm{nm}$, $\lambda_\textrm{D} = 9.61\:\textrm{nm}$, $\hat{U} = 1.25\:\textrm{V}$) can be obtained. Hence, if quasi-stationary vortex flow is desired, two counter-propagating waves instead of two standing waves should be used.

It is known that the standard theory of ICEOF often overpredicts flow velocities obtained experimentally \cite{Sugioka:PRE2016}, while ACEOF experiments conducted at high frequencies and voltages comparable to those used in this study indicate that one may observe flow reversals that are not captured by the standard Poisson-Nernst-Planck (PNP) framework underlying this study \cite{Bazant:AdvCollIntScie2009}. The reasons for these discrepancies are still in the focus of active research. Possible explanations include phase-delay \cite{Sugioka:PRE2016} and excluded-volume effects \cite{Bazant:AdvCollIntScie2009}. While the former effect is included in our work, for the latter, the ion distributions under excluded-volume effects in non-equilibrium ICEOF scenarios are frequently postulated to follow a Fermi-Dirac statistics. Since the charge density would still be a function of the electric potential as the only variable parameter, describing the ion density in a non-equilibrium situation with an equilibrium distribution function appears to be questionable. According to expression (\ref{Eq:EOP}), the EOP is zero for such a case, i.e., no flow, including flow reversals, can develop. Hence, the explanation of flow reversal by excluded-volume effects still requires further clarification. In addition, the electrode and counter-electrode used in ACEOF experiments exhibiting flow reversals typically differ in size \cite{Bazant:AdvCollIntScie2009}. This is not an issue in this work. Finally, other authors have suggested that the flow reversal is due to the combined effects of non-identical mobilities of the involved ion species and Faradaic currents at the electrodes \cite{Gonzalez:PRE2010}. While these effects are important and relevant, we think that further detailing the model in an attempt to capture these issues would go beyond the basic scope of this study, namely to address electrokinetic effects and the affiliated fundamental flow physics induced by SAWs that cause non-negligible EOF particularly in nanochannels. For this reason, we leave the problem of flow reversals to future studies.

Figure \ref{Fig:TVLSAW_uax_zeta_sweep} shows the time-averaged axial velocity $\langle u \rangle_\textrm{ax}$, averaged over the channel cross section, as a function of the surface potential amplitude $\hat{U}$ and with $f_\textrm{SAW}$ as a parameter. All cases follow the quadratic dependence as one may expect from Eq. (\ref{Eq:TWACEOF_velmax}), with the largest velocities developing at $f_\textrm{SAW} = 10\:\textrm{MHz}$. This plot is another demonstration that the physical mechanism underlying SAW-EOF is similar to that of ACEOF. It also suggests that for even higher potential amplitudes - not uncommon in practical realizations of SAWs - significantly higher flow velocities than those reported in this work may be induced.
\begin{figure}
	\centerline{\includegraphics[width=6.5cm]{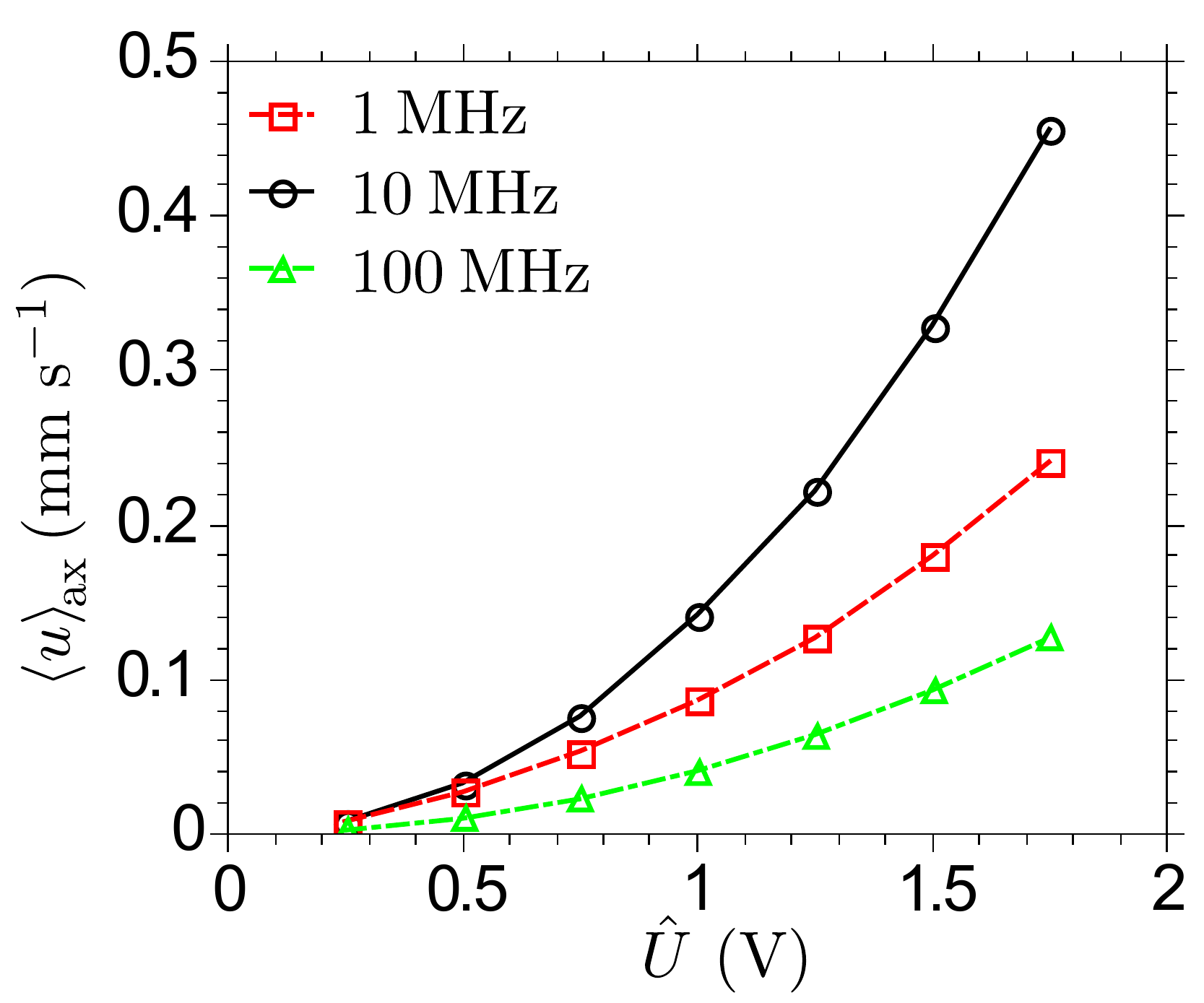}}
	\caption{Time-averaged axial velocity $\langle u \rangle_\textrm{ax}$ as a function of the surface potential amplitude $\hat{U}$ and with $f_\textrm{SAW}$ as parameter, while $\Delta \varphi = \pi$, $h = 480\:\textrm{nm}$, $\lambda_\textrm{D} = 9.61\:\textrm{nm}$, and $n_0 = 1\:\textrm{mM}$. The lines connecting the data points are guides to the eye.}
\label{Fig:TVLSAW_uax_zeta_sweep}
\end{figure}

\section{Conclusions}

In this work, the electroosmotic flow induced in an aqueous solution by surface acoustic waves standing or traveling along (piezo-active) walls of a narrow parallel-plate channel has been investigated numerically. The physical mechanism is similar to traveling wave ACEOF, but with no electrodes at the channel walls. For standing waves, it was seen that vortex pairs develop that change their sense of rotation four times per cycle. The frequent flow reversal implies a noticeable but small net velocity of ${\cal O}(10^1)\:\mu\textrm{m s}^{-1}$ that is maximized when the waves at the opposing channel walls have an identical frequency but a phase shift of $90^\circ$. For traveling waves, it was found that similar vortex pairs develop that, however, do not change their sense of rotation. Instead, they move along the channel, leading to a net flow that scales quadratically with the voltage amplitude. Due to the absence of periodic reversals of the vortex pairs, higher net velocities can be generated that can be of ${\cal O}(10^{-1})\:\textrm{mm s}^{-1}$. For frequencies of the same order of magnitude as the inverse of the nominal RC-time of the electric double layer, transport is maximized if acoustic waves of identical frequency but phase-shifted by $180^\circ$ are imposed on both channel walls facing each other. This maximizes the electric current between the oppositely charged Debye layers on these walls. Conventionally, for mm-sized flow domains, surface acoustic waves may generate flow velocities of ${\cal O}(10^{1})\:\textrm{cm s}^{-1}$ by means of acoustic streaming. However, this actuation method is inefficient for narrow nanometer-wide channels, so that for such cases the EOF described in this work may be the dominant mode of transport. Given that it does not require any elaborate wiring inside of the channel, it may be an interesting approach to drive liquids through narrow confinements.

\section{Acknowledgment}
Financial support by the German Research Foundation (DFG) through Grant No. HA 2696/42-1 is gratefully acknowledged.

\bibliography{references}

\begin{thebibliography}{69}%
\makeatletter
\providecommand \@ifxundefined [1]{%
 \@ifx{#1\undefined}
}%
\providecommand \@ifnum [1]{%
 \ifnum #1\expandafter \@firstoftwo
 \else \expandafter \@secondoftwo
 \fi
}%
\providecommand \@ifx [1]{%
 \ifx #1\expandafter \@firstoftwo
 \else \expandafter \@secondoftwo
 \fi
}%
\providecommand \natexlab [1]{#1}%
\providecommand \enquote  [1]{``#1''}%
\providecommand \bibnamefont  [1]{#1}%
\providecommand \bibfnamefont [1]{#1}%
\providecommand \citenamefont [1]{#1}%
\providecommand \href@noop [0]{\@secondoftwo}%
\providecommand \href [0]{\begingroup \@sanitize@url \@href}%
\providecommand \@href[1]{\@@startlink{#1}\@@href}%
\providecommand \@@href[1]{\endgroup#1\@@endlink}%
\providecommand \@sanitize@url [0]{\catcode `\\12\catcode `\$12\catcode
  `\&12\catcode `\#12\catcode `\^12\catcode `\_12\catcode `\%12\relax}%
\providecommand \@@startlink[1]{}%
\providecommand \@@endlink[0]{}%
\providecommand \url  [0]{\begingroup\@sanitize@url \@url }%
\providecommand \@url [1]{\endgroup\@href {#1}{\urlprefix }}%
\providecommand \urlprefix  [0]{URL }%
\providecommand \Eprint [0]{\href }%
\providecommand \doibase [0]{https://doi.org/}%
\providecommand \selectlanguage [0]{\@gobble}%
\providecommand \bibinfo  [0]{\@secondoftwo}%
\providecommand \bibfield  [0]{\@secondoftwo}%
\providecommand \translation [1]{[#1]}%
\providecommand \BibitemOpen [0]{}%
\providecommand \bibitemStop [0]{}%
\providecommand \bibitemNoStop [0]{.\EOS\space}%
\providecommand \EOS [0]{\spacefactor3000\relax}%
\providecommand \BibitemShut  [1]{\csname bibitem#1\endcsname}%
\let\auto@bib@innerbib\@empty
\bibitem [{\citenamefont {Franke}\ \emph {et~al.}(2009)\citenamefont {Franke},
  \citenamefont {Abate}, \citenamefont {Weitz},\ and\ \citenamefont
  {Wixforth}}]{Franke:LabChip2009}%
  \BibitemOpen
  \bibfield  {author} {\bibinfo {author} {\bibfnamefont {T.}~\bibnamefont
  {Franke}}, \bibinfo {author} {\bibfnamefont {A.~R.}\ \bibnamefont {Abate}},
  \bibinfo {author} {\bibfnamefont {D.~A.}\ \bibnamefont {Weitz}},\ and\
  \bibinfo {author} {\bibfnamefont {A.}~\bibnamefont {Wixforth}},\ }\bibfield
  {title} {\bibinfo {title} {Surface acoustic wave ($\textrm{SAW}$) directed
  droplet flow in microfluidics for $\textrm{PDMS}$ devices},\ }\href@noop {}
  {\bibfield  {journal} {\bibinfo  {journal} {Lab Chip}\ }\textbf {\bibinfo
  {volume} {9}},\ \bibinfo {pages} {2625} (\bibinfo {year} {2009})}\BibitemShut
  {NoStop}%
\bibitem [{\citenamefont {Dentry}\ \emph {et~al.}(2014)\citenamefont {Dentry},
  \citenamefont {Yeo},\ and\ \citenamefont {Friend}}]{Dentry:PRE2014}%
  \BibitemOpen
  \bibfield  {author} {\bibinfo {author} {\bibfnamefont {M.~B.}\ \bibnamefont
  {Dentry}}, \bibinfo {author} {\bibfnamefont {L.~Y.}\ \bibnamefont {Yeo}},\
  and\ \bibinfo {author} {\bibfnamefont {J.~R.}\ \bibnamefont {Friend}},\
  }\bibfield  {title} {\bibinfo {title} {Frequency effects on the scale and
  behavior of acoustic streaming},\ }\href@noop {} {\bibfield  {journal}
  {\bibinfo  {journal} {Phys.~Rev.~E}\ }\textbf {\bibinfo {volume} {89}},\
  \bibinfo {pages} {013203} (\bibinfo {year} {2014})}\BibitemShut {NoStop}%
\bibitem [{\citenamefont {Wiklund}\ \emph {et~al.}(2012)\citenamefont
  {Wiklund}, \citenamefont {Green},\ and\ \citenamefont
  {Ohlin}}]{Wiklund:LabChip2012}%
  \BibitemOpen
  \bibfield  {author} {\bibinfo {author} {\bibfnamefont {M.}~\bibnamefont
  {Wiklund}}, \bibinfo {author} {\bibfnamefont {R.}~\bibnamefont {Green}},\
  and\ \bibinfo {author} {\bibfnamefont {M.}~\bibnamefont {Ohlin}},\ }\bibfield
   {title} {\bibinfo {title} {Acoustofluidics 14: Applications of acoustic
  streaming in microfluidic devices},\ }\href@noop {} {\bibfield  {journal}
  {\bibinfo  {journal} {Lab Chip}\ }\textbf {\bibinfo {volume} {12}},\ \bibinfo
  {pages} {2438} (\bibinfo {year} {2012})}\BibitemShut {NoStop}%
\bibitem [{\citenamefont {Sadhal}(2012)}]{Sadhal:LabChip2012}%
  \BibitemOpen
  \bibfield  {author} {\bibinfo {author} {\bibfnamefont {S.~S.}\ \bibnamefont
  {Sadhal}},\ }\bibfield  {title} {\bibinfo {title} {Acoustofluidics 13:
  Analysis of acoustic streaming by perturbation methods},\ }\href@noop {}
  {\bibfield  {journal} {\bibinfo  {journal} {Lab~Chip}\ }\textbf {\bibinfo
  {volume} {12}},\ \bibinfo {pages} {2292} (\bibinfo {year}
  {2012})}\BibitemShut {NoStop}%
\bibitem [{\citenamefont {Wu}(2018)}]{Wu:Fluids2018}%
  \BibitemOpen
  \bibfield  {author} {\bibinfo {author} {\bibfnamefont {J.}~\bibnamefont
  {Wu}},\ }\bibfield  {title} {\bibinfo {title} {Acoustic streaming and its
  applications},\ }\href@noop {} {\bibfield  {journal} {\bibinfo  {journal}
  {Fluids}\ }\textbf {\bibinfo {volume} {3}},\ \bibinfo {pages} {108} (\bibinfo
  {year} {2018})}\BibitemShut {NoStop}%
\bibitem [{\citenamefont {Eckart}(1947)}]{Eckart:PhysRev1947}%
  \BibitemOpen
  \bibfield  {author} {\bibinfo {author} {\bibfnamefont {C.}~\bibnamefont
  {Eckart}},\ }\bibfield  {title} {\bibinfo {title} {Vortices and streams
  caused by sound waves},\ }\href@noop {} {\bibfield  {journal} {\bibinfo
  {journal} {Phys.~Rev.}\ }\textbf {\bibinfo {volume} {73}},\ \bibinfo {pages}
  {68} (\bibinfo {year} {1947})}\BibitemShut {NoStop}%
\bibitem [{\citenamefont {Riley}(2001)}]{Riley:AnnRevFluidMech2001}%
  \BibitemOpen
  \bibfield  {author} {\bibinfo {author} {\bibfnamefont {N.}~\bibnamefont
  {Riley}},\ }\bibfield  {title} {\bibinfo {title} {Steady streaming},\
  }\href@noop {} {\bibfield  {journal} {\bibinfo  {journal} {Annu. Rev. Fluid
  Mech.}\ }\textbf {\bibinfo {volume} {33}},\ \bibinfo {pages} {43} (\bibinfo
  {year} {2001})}\BibitemShut {NoStop}%
\bibitem [{\citenamefont {Nybord}(1958)}]{Nyborg:JAcoustSocAm1958}%
  \BibitemOpen
  \bibfield  {author} {\bibinfo {author} {\bibfnamefont {W.~L.}\ \bibnamefont
  {Nybord}},\ }\bibfield  {title} {\bibinfo {title} {Acoustic streaming near a
  boundary},\ }\href@noop {} {\bibfield  {journal} {\bibinfo  {journal}
  {J.~Acoust.~Soc.~Am.}\ }\textbf {\bibinfo {volume} {30}},\ \bibinfo {pages}
  {329} (\bibinfo {year} {1958})}\BibitemShut {NoStop}%
\bibitem [{\citenamefont {Stokes}(1845)}]{Stokes:TCPS1845}%
  \BibitemOpen
  \bibfield  {author} {\bibinfo {author} {\bibfnamefont {G.~G.}\ \bibnamefont
  {Stokes}},\ }\bibfield  {title} {\bibinfo {title} {On the theories of the
  internal friction in fluids in motion, and of the equilibrium and motion of
  elastic solids},\ }\href@noop {} {\bibfield  {journal} {\bibinfo  {journal}
  {Trans.~Cambridge Philos.~Soc.}\ }\textbf {\bibinfo {volume} {8}},\ \bibinfo
  {pages} {287} (\bibinfo {year} {1845})}\BibitemShut {NoStop}%
\bibitem [{\citenamefont {Kurosawa}\ \emph {et~al.}(1995)\citenamefont
  {Kurosawa}, \citenamefont {Watanabe}, \citenamefont {Futami},\ and\
  \citenamefont {Higuchi}}]{Kurosawa:SensAct1995}%
  \BibitemOpen
  \bibfield  {author} {\bibinfo {author} {\bibfnamefont {M.}~\bibnamefont
  {Kurosawa}}, \bibinfo {author} {\bibfnamefont {T.}~\bibnamefont {Watanabe}},
  \bibinfo {author} {\bibfnamefont {A.}~\bibnamefont {Futami}},\ and\ \bibinfo
  {author} {\bibfnamefont {T.}~\bibnamefont {Higuchi}},\ }\bibfield  {title}
  {\bibinfo {title} {Surface acoustic wave atomizer},\ }\href@noop {}
  {\bibfield  {journal} {\bibinfo  {journal} {Sens.~Actuators A}\ }\textbf
  {\bibinfo {volume} {50}},\ \bibinfo {pages} {69} (\bibinfo {year}
  {1995})}\BibitemShut {NoStop}%
\bibitem [{\citenamefont {Qi}\ \emph {et~al.}(2008)\citenamefont {Qi},
  \citenamefont {Yeo},\ and\ \citenamefont {Friend}}]{Qi:PoF2008}%
  \BibitemOpen
  \bibfield  {author} {\bibinfo {author} {\bibfnamefont {A.}~\bibnamefont
  {Qi}}, \bibinfo {author} {\bibfnamefont {L.~Y.}\ \bibnamefont {Yeo}},\ and\
  \bibinfo {author} {\bibfnamefont {J.~R.}\ \bibnamefont {Friend}},\ }\bibfield
   {title} {\bibinfo {title} {Interfacial destabilization and atomization
  driven by surface acoustic waves},\ }\href@noop {} {\bibfield  {journal}
  {\bibinfo  {journal} {Phys.~Fluids}\ }\textbf {\bibinfo {volume} {20}},\
  \bibinfo {pages} {074103} (\bibinfo {year} {2008})}\BibitemShut {NoStop}%
\bibitem [{\citenamefont {Yeo}\ and\ \citenamefont
  {Friend}(2014)}]{Yeo:AnnRevFluidMech2014}%
  \BibitemOpen
  \bibfield  {author} {\bibinfo {author} {\bibfnamefont {L.~Y.}\ \bibnamefont
  {Yeo}}\ and\ \bibinfo {author} {\bibfnamefont {J.~R.}\ \bibnamefont
  {Friend}},\ }\bibfield  {title} {\bibinfo {title} {Surface acoustic wave
  microfluidics},\ }\href@noop {} {\bibfield  {journal} {\bibinfo  {journal}
  {Annu.~Rev.~Fluid Mech.}\ }\textbf {\bibinfo {volume} {46}},\ \bibinfo
  {pages} {379} (\bibinfo {year} {2014})}\BibitemShut {NoStop}%
\bibitem [{\citenamefont {Beyssen}\ \emph {et~al.}(2006)\citenamefont
  {Beyssen}, \citenamefont {Brizoual}, \citenamefont {Elmazria},\ and\
  \citenamefont {Alnot}}]{Beyssen:SensActB2006}%
  \BibitemOpen
  \bibfield  {author} {\bibinfo {author} {\bibfnamefont {D.}~\bibnamefont
  {Beyssen}}, \bibinfo {author} {\bibfnamefont {L.~L.}\ \bibnamefont
  {Brizoual}}, \bibinfo {author} {\bibfnamefont {O.}~\bibnamefont {Elmazria}},\
  and\ \bibinfo {author} {\bibfnamefont {P.}~\bibnamefont {Alnot}},\ }\bibfield
   {title} {\bibinfo {title} {Microfluidic device based on surface acoustic
  wave},\ }\href@noop {} {\bibfield  {journal} {\bibinfo  {journal}
  {Sens.~Actuators B}\ }\textbf {\bibinfo {volume} {118}},\ \bibinfo {pages}
  {380} (\bibinfo {year} {2006})}\BibitemShut {NoStop}%
\bibitem [{\citenamefont {Yeo}\ and\ \citenamefont
  {Friend}(2009)}]{Yeo:Biomicrofluidics2009}%
  \BibitemOpen
  \bibfield  {author} {\bibinfo {author} {\bibfnamefont {L.~Y.}\ \bibnamefont
  {Yeo}}\ and\ \bibinfo {author} {\bibfnamefont {J.~R.}\ \bibnamefont
  {Friend}},\ }\bibfield  {title} {\bibinfo {title} {Ultrafast microfluidics
  using surface acoustic waves},\ }\href@noop {} {\bibfield  {journal}
  {\bibinfo  {journal} {Biomicrofluidics}\ }\textbf {\bibinfo {volume} {3}},\
  \bibinfo {pages} {012002} (\bibinfo {year} {2009})}\BibitemShut {NoStop}%
\bibitem [{\citenamefont {Lei}\ \emph {et~al.}(2018)\citenamefont {Lei},
  \citenamefont {Hill}, \citenamefont {de~Le\'on~Albarr\'an},\ and\
  \citenamefont {Glynne-Jones}}]{Lei:MANO2018}%
  \BibitemOpen
  \bibfield  {author} {\bibinfo {author} {\bibfnamefont {J.}~\bibnamefont
  {Lei}}, \bibinfo {author} {\bibfnamefont {M.}~\bibnamefont {Hill}}, \bibinfo
  {author} {\bibfnamefont {C.~P.}\ \bibnamefont {de~Le\'on~Albarr\'an}},\ and\
  \bibinfo {author} {\bibfnamefont {P.}~\bibnamefont {Glynne-Jones}},\
  }\bibfield  {title} {\bibinfo {title} {Effects of micron scale surface
  profiles on acoustic streaming},\ }\href@noop {} {\bibfield  {journal}
  {\bibinfo  {journal} {Microfluid.~Nanofluid.}\ }\textbf {\bibinfo {volume}
  {22}},\ \bibinfo {pages} {140} (\bibinfo {year} {2018})}\BibitemShut
  {NoStop}%
\bibitem [{\citenamefont {Tan}\ \emph {et~al.}(2009)\citenamefont {Tan},
  \citenamefont {Yeo},\ and\ \citenamefont {Friend}}]{Tan:EPL2009}%
  \BibitemOpen
  \bibfield  {author} {\bibinfo {author} {\bibfnamefont {M.~K.}\ \bibnamefont
  {Tan}}, \bibinfo {author} {\bibfnamefont {L.~Y.}\ \bibnamefont {Yeo}},\ and\
  \bibinfo {author} {\bibfnamefont {J.~R.}\ \bibnamefont {Friend}},\ }\bibfield
   {title} {\bibinfo {title} {Rapid fluid flow and mixing induced in
  microchannels using surface acoustic waves},\ }\href@noop {} {\bibfield
  {journal} {\bibinfo  {journal} {EPL}\ }\textbf {\bibinfo {volume} {87}},\
  \bibinfo {pages} {47003} (\bibinfo {year} {2009})}\BibitemShut {NoStop}%
\bibitem [{\citenamefont {Schmid}\ \emph {et~al.}(2012)\citenamefont {Schmid},
  \citenamefont {Wixforth}, \citenamefont {Weitz},\ and\ \citenamefont
  {Franke}}]{Schmid:MANO2012}%
  \BibitemOpen
  \bibfield  {author} {\bibinfo {author} {\bibfnamefont {L.}~\bibnamefont
  {Schmid}}, \bibinfo {author} {\bibfnamefont {A.}~\bibnamefont {Wixforth}},
  \bibinfo {author} {\bibfnamefont {D.~A.}\ \bibnamefont {Weitz}},\ and\
  \bibinfo {author} {\bibfnamefont {T.}~\bibnamefont {Franke}},\ }\bibfield
  {title} {\bibinfo {title} {Novel surface acoustic wave
  ($\textrm{SAW}$)-driven closed $\textrm{PDMS}$ flow chamber},\ }\href@noop {}
  {\bibfield  {journal} {\bibinfo  {journal} {Microfluid.~Nanofluid.}\ }\textbf
  {\bibinfo {volume} {12}},\ \bibinfo {pages} {229} (\bibinfo {year}
  {2012})}\BibitemShut {NoStop}%
\bibitem [{\citenamefont {Kiebert}\ \emph {et~al.}(2017)\citenamefont
  {Kiebert}, \citenamefont {Wege}, \citenamefont {Massing}, \citenamefont
  {K\"onig}, \citenamefont {Cierpka}, \citenamefont {Wesera},\ and\
  \citenamefont {Schmidt}}]{Kiebert:LabChip2017}%
  \BibitemOpen
  \bibfield  {author} {\bibinfo {author} {\bibfnamefont {F.}~\bibnamefont
  {Kiebert}}, \bibinfo {author} {\bibfnamefont {S.}~\bibnamefont {Wege}},
  \bibinfo {author} {\bibfnamefont {J.}~\bibnamefont {Massing}}, \bibinfo
  {author} {\bibfnamefont {J.}~\bibnamefont {K\"onig}}, \bibinfo {author}
  {\bibfnamefont {C.}~\bibnamefont {Cierpka}}, \bibinfo {author} {\bibfnamefont
  {R.}~\bibnamefont {Wesera}},\ and\ \bibinfo {author} {\bibfnamefont
  {H.}~\bibnamefont {Schmidt}},\ }\bibfield  {title} {\bibinfo {title}
  {3$\textrm{D}$ measurement and simulation of surface acoustic wave driven
  fluid motion: a comparison},\ }\href@noop {} {\bibfield  {journal} {\bibinfo
  {journal} {Lab Chip}\ }\textbf {\bibinfo {volume} {17}},\ \bibinfo {pages}
  {2104} (\bibinfo {year} {2017})}\BibitemShut {NoStop}%
\bibitem [{\citenamefont {Hags\"ater}\ \emph {et~al.}(2007)\citenamefont
  {Hags\"ater}, \citenamefont {Jensen}, \citenamefont {Bruus},\ and\
  \citenamefont {Kutter}}]{Hagsaeter:LabChip2007}%
  \BibitemOpen
  \bibfield  {author} {\bibinfo {author} {\bibfnamefont {S.~M.}\ \bibnamefont
  {Hags\"ater}}, \bibinfo {author} {\bibfnamefont {T.~G.}\ \bibnamefont
  {Jensen}}, \bibinfo {author} {\bibfnamefont {H.}~\bibnamefont {Bruus}},\ and\
  \bibinfo {author} {\bibfnamefont {J.~P.}\ \bibnamefont {Kutter}},\ }\bibfield
   {title} {\bibinfo {title} {Acoustic resonances in microfluidic chips:
  full-image micro-$\textrm{PIV}$ experiments and numerical simulations},\
  }\href@noop {} {\bibfield  {journal} {\bibinfo  {journal} {Lab Chip}\
  }\textbf {\bibinfo {volume} {7}},\ \bibinfo {pages} {1336} (\bibinfo {year}
  {2007})}\BibitemShut {NoStop}%
\bibitem [{\citenamefont {Cecchini}\ \emph {et~al.}(2008)\citenamefont
  {Cecchini}, \citenamefont {Girardo}, \citenamefont {Pisignano}, \citenamefont
  {Cingolani},\ and\ \citenamefont {Beltram}}]{Cecchini:APL2008}%
  \BibitemOpen
  \bibfield  {author} {\bibinfo {author} {\bibfnamefont {M.}~\bibnamefont
  {Cecchini}}, \bibinfo {author} {\bibfnamefont {S.}~\bibnamefont {Girardo}},
  \bibinfo {author} {\bibfnamefont {D.}~\bibnamefont {Pisignano}}, \bibinfo
  {author} {\bibfnamefont {R.}~\bibnamefont {Cingolani}},\ and\ \bibinfo
  {author} {\bibfnamefont {F.}~\bibnamefont {Beltram}},\ }\bibfield  {title}
  {\bibinfo {title} {Acoustic-counterflow microfluidics by surface acoustic
  waves},\ }\href@noop {} {\bibfield  {journal} {\bibinfo  {journal}
  {Appl.~Phys.~Lett.}\ }\textbf {\bibinfo {volume} {92}},\ \bibinfo {pages}
  {104103} (\bibinfo {year} {2008})}\BibitemShut {NoStop}%
\bibitem [{\citenamefont {Xie}\ and\ \citenamefont {Cao}(2017)}]{Xie:MANO2017}%
  \BibitemOpen
  \bibfield  {author} {\bibinfo {author} {\bibfnamefont {J.~F.}\ \bibnamefont
  {Xie}}\ and\ \bibinfo {author} {\bibfnamefont {B.~Y.}\ \bibnamefont {Cao}},\
  }\bibfield  {title} {\bibinfo {title} {Fast nanofluidics by travelling
  surface waves},\ }\href@noop {} {\bibfield  {journal} {\bibinfo  {journal}
  {Microfluid.~Nanofluid.}\ }\textbf {\bibinfo {volume} {21}},\ \bibinfo
  {pages} {111} (\bibinfo {year} {2017})}\BibitemShut {NoStop}%
\bibitem [{\citenamefont {Tan}\ and\ \citenamefont {Yeo}(2018)}]{Tan:PRF2018}%
  \BibitemOpen
  \bibfield  {author} {\bibinfo {author} {\bibfnamefont {M.~K.}\ \bibnamefont
  {Tan}}\ and\ \bibinfo {author} {\bibfnamefont {L.~Y.}\ \bibnamefont {Yeo}},\
  }\bibfield  {title} {\bibinfo {title} {Hybrid finite-difference/lattice
  boltzmann simulations of microchannel and nanochannel acoustic streaming
  driven by surface acoustic waves},\ }\href@noop {} {\bibfield  {journal}
  {\bibinfo  {journal} {Phys.~Rev.~Fluids}\ }\textbf {\bibinfo {volume} {3}},\
  \bibinfo {pages} {044202} (\bibinfo {year} {2018})}\BibitemShut {NoStop}%
\bibitem [{\citenamefont {Jakubik}(2014)}]{Jakubik:SensActB2014}%
  \BibitemOpen
  \bibfield  {author} {\bibinfo {author} {\bibfnamefont {W.}~\bibnamefont
  {Jakubik}},\ }\bibfield  {title} {\bibinfo {title} {Elemental theory of a
  $\textrm{SAW}$ gas sensor based on electrical conductivity changes in
  bi-layer nanostructures},\ }\href@noop {} {\bibfield  {journal} {\bibinfo
  {journal} {Sens.~Actuators B}\ }\textbf {\bibinfo {volume} {203}},\ \bibinfo
  {pages} {511} (\bibinfo {year} {2014})}\BibitemShut {NoStop}%
\bibitem [{\citenamefont {Huang}\ \emph {et~al.}(2020)\citenamefont {Huang},
  \citenamefont {Liu}, \citenamefont {Manor}, \citenamefont {Liu},\ and\
  \citenamefont {Friend}}]{Huang:AdvMat2020}%
  \BibitemOpen
  \bibfield  {author} {\bibinfo {author} {\bibfnamefont {A.}~\bibnamefont
  {Huang}}, \bibinfo {author} {\bibfnamefont {H.}~\bibnamefont {Liu}}, \bibinfo
  {author} {\bibfnamefont {O.}~\bibnamefont {Manor}}, \bibinfo {author}
  {\bibfnamefont {P.}~\bibnamefont {Liu}},\ and\ \bibinfo {author}
  {\bibfnamefont {J.}~\bibnamefont {Friend}},\ }\bibfield  {title} {\bibinfo
  {title} {Enabling rapid charging lithium metal batteries via surface acoustic
  wave-driven electrolyte flow},\ }\href@noop {} {\bibfield  {journal}
  {\bibinfo  {journal} {Adv.~Mater.}\ }\textbf {\bibinfo {volume} {32}},\
  \bibinfo {pages} {1907516} (\bibinfo {year} {2020})}\BibitemShut {NoStop}%
\bibitem [{\citenamefont {Josse}\ \emph {et~al.}(1992)\citenamefont {Josse},
  \citenamefont {Shana}, \citenamefont {Hawortha}, \citenamefont {Liew},\ and\
  \citenamefont {Grunze}}]{Josse:SensActB1992}%
  \BibitemOpen
  \bibfield  {author} {\bibinfo {author} {\bibfnamefont {F.}~\bibnamefont
  {Josse}}, \bibinfo {author} {\bibfnamefont {Z.~A.}\ \bibnamefont {Shana}},
  \bibinfo {author} {\bibfnamefont {D.~T.}\ \bibnamefont {Hawortha}}, \bibinfo
  {author} {\bibfnamefont {S.}~\bibnamefont {Liew}},\ and\ \bibinfo {author}
  {\bibfnamefont {M.}~\bibnamefont {Grunze}},\ }\bibfield  {title} {\bibinfo
  {title} {On the use of $\textrm{ZX}$-$\textrm{LiNbO}_3$, acoustic plate mode
  devices as detectors for dilute electrolytes},\ }\href@noop {} {\bibfield
  {journal} {\bibinfo  {journal} {Sens.~Act.~B}\ }\textbf {\bibinfo {volume}
  {9}},\ \bibinfo {pages} {97} (\bibinfo {year} {1992})}\BibitemShut {NoStop}%
\bibitem [{\citenamefont {Son}(2004)}]{Son:ElectrochemComm2004}%
  \BibitemOpen
  \bibfield  {author} {\bibinfo {author} {\bibfnamefont {J.}~\bibnamefont
  {Son}},\ }\bibfield  {title} {\bibinfo {title} {Novel electrode material for
  li ion battery based on polycrystalline
  $\textrm{L}$i$\textrm{N}$b$\textrm{O}_3$},\ }\href@noop {} {\bibfield
  {journal} {\bibinfo  {journal} {Electrochem.~Commun.}\ }\textbf {\bibinfo
  {volume} {6}},\ \bibinfo {pages} {990} (\bibinfo {year} {2004})}\BibitemShut
  {NoStop}%
\bibitem [{\citenamefont {Liu}\ \emph {et~al.}(2014)\citenamefont {Liu},
  \citenamefont {Shakir},\ and\ \citenamefont {Kang}}]{Lui:MatLett2014}%
  \BibitemOpen
  \bibfield  {author} {\bibinfo {author} {\bibfnamefont {J.}~\bibnamefont
  {Liu}}, \bibinfo {author} {\bibfnamefont {I.}~\bibnamefont {Shakir}},\ and\
  \bibinfo {author} {\bibfnamefont {D.~J.}\ \bibnamefont {Kang}},\ }\bibfield
  {title} {\bibinfo {title} {Lithium niobate nanoflakes as electrodes for
  highly stable electrochemical supercapacitor devices},\ }\href@noop {}
  {\bibfield  {journal} {\bibinfo  {journal} {Mater.~Lett.}\ }\textbf {\bibinfo
  {volume} {119}},\ \bibinfo {pages} {84} (\bibinfo {year} {2014})}\BibitemShut
  {NoStop}%
\bibitem [{\citenamefont {Grilli}\ \emph {et~al.}(2008)\citenamefont {Grilli},
  \citenamefont {Miccio}, \citenamefont {Vespini}, \citenamefont {Finizio},
  \citenamefont {Nicola},\ and\ \citenamefont
  {Ferraro}}]{Grilli:OpticsExpr2008}%
  \BibitemOpen
  \bibfield  {author} {\bibinfo {author} {\bibfnamefont {S.}~\bibnamefont
  {Grilli}}, \bibinfo {author} {\bibfnamefont {L.}~\bibnamefont {Miccio}},
  \bibinfo {author} {\bibfnamefont {V.}~\bibnamefont {Vespini}}, \bibinfo
  {author} {\bibfnamefont {A.}~\bibnamefont {Finizio}}, \bibinfo {author}
  {\bibfnamefont {S.~D.}\ \bibnamefont {Nicola}},\ and\ \bibinfo {author}
  {\bibfnamefont {P.}~\bibnamefont {Ferraro}},\ }\bibfield  {title} {\bibinfo
  {title} {Liquid micro-lens array activated by selective electrowetting on
  lithium niobate substrates},\ }\href@noop {} {\bibfield  {journal} {\bibinfo
  {journal} {Opt.~Express}\ }\textbf {\bibinfo {volume} {16}},\ \bibinfo
  {pages} {8084} (\bibinfo {year} {2008})}\BibitemShut {NoStop}%
\bibitem [{\citenamefont {Fitts}(1962)}]{Fitts:McGrawHill1962}%
  \BibitemOpen
  \bibfield  {author} {\bibinfo {author} {\bibfnamefont {D.~D.}\ \bibnamefont
  {Fitts}},\ }in\ \href@noop {} {\emph {\bibinfo {booktitle} {Non-equilibrium
  thermodynamics}}}\ (\bibinfo  {publisher} {McGraw-Hills, New York},\ \bibinfo
  {year} {1962})\BibitemShut {NoStop}%
\bibitem [{\citenamefont {Squires}\ and\ \citenamefont
  {Bazant}(2004)}]{Squires:JFM2004}%
  \BibitemOpen
  \bibfield  {author} {\bibinfo {author} {\bibfnamefont {T.~M.}\ \bibnamefont
  {Squires}}\ and\ \bibinfo {author} {\bibfnamefont {M.~Z.}\ \bibnamefont
  {Bazant}},\ }\bibfield  {title} {\bibinfo {title} {Induced-charge
  electro-osmosis},\ }\href@noop {} {\bibfield  {journal} {\bibinfo  {journal}
  {J.~Fluid Mech.}\ }\textbf {\bibinfo {volume} {509}},\ \bibinfo {pages} {217}
  (\bibinfo {year} {2004})}\BibitemShut {NoStop}%
\bibitem [{\citenamefont {Suh}\ and\ \citenamefont
  {Kang}(2010)}]{Suh:Micromachines2010}%
  \BibitemOpen
  \bibfield  {author} {\bibinfo {author} {\bibfnamefont {Y.~K.}\ \bibnamefont
  {Suh}}\ and\ \bibinfo {author} {\bibfnamefont {S.}~\bibnamefont {Kang}},\
  }\bibfield  {title} {\bibinfo {title} {A review on mixing in microfluidics},\
  }\href@noop {} {\bibfield  {journal} {\bibinfo  {journal} {Micromachines}\
  }\textbf {\bibinfo {volume} {1}},\ \bibinfo {pages} {82} (\bibinfo {year}
  {2010})}\BibitemShut {NoStop}%
\bibitem [{\citenamefont {Luka}\ \emph {et~al.}(2015)\citenamefont {Luka},
  \citenamefont {Ahmadi}, \citenamefont {Najjaran}, \citenamefont {Alocilja},
  \citenamefont {DeRosa}, \citenamefont {Wolthers}, \citenamefont {Malki},
  \citenamefont {Aziz}, \citenamefont {Althani},\ and\ \citenamefont
  {Hoorfar}}]{Luka:Sensors2015}%
  \BibitemOpen
  \bibfield  {author} {\bibinfo {author} {\bibfnamefont {G.}~\bibnamefont
  {Luka}}, \bibinfo {author} {\bibfnamefont {A.}~\bibnamefont {Ahmadi}},
  \bibinfo {author} {\bibfnamefont {H.}~\bibnamefont {Najjaran}}, \bibinfo
  {author} {\bibfnamefont {E.}~\bibnamefont {Alocilja}}, \bibinfo {author}
  {\bibfnamefont {M.}~\bibnamefont {DeRosa}}, \bibinfo {author} {\bibfnamefont
  {K.}~\bibnamefont {Wolthers}}, \bibinfo {author} {\bibfnamefont
  {A.}~\bibnamefont {Malki}}, \bibinfo {author} {\bibfnamefont
  {H.}~\bibnamefont {Aziz}}, \bibinfo {author} {\bibfnamefont {A.}~\bibnamefont
  {Althani}},\ and\ \bibinfo {author} {\bibfnamefont {M.}~\bibnamefont
  {Hoorfar}},\ }\bibfield  {title} {\bibinfo {title} {Microfluidics integrated
  biosensors: A leading technology towards lab-on-a-chip and sensing
  applications},\ }\href@noop {} {\bibfield  {journal} {\bibinfo  {journal}
  {Sensors}\ }\textbf {\bibinfo {volume} {15}},\ \bibinfo {pages} {30011}
  (\bibinfo {year} {2015})}\BibitemShut {NoStop}%
\bibitem [{\citenamefont {Olanrewaju}\ \emph {et~al.}(2018)\citenamefont
  {Olanrewaju}, \citenamefont {Beaugrand}, \citenamefont {Yafia},\ and\
  \citenamefont {Juncker}}]{Olanrewaju:LabChip2018}%
  \BibitemOpen
  \bibfield  {author} {\bibinfo {author} {\bibfnamefont {A.}~\bibnamefont
  {Olanrewaju}}, \bibinfo {author} {\bibfnamefont {M.}~\bibnamefont
  {Beaugrand}}, \bibinfo {author} {\bibfnamefont {M.}~\bibnamefont {Yafia}},\
  and\ \bibinfo {author} {\bibfnamefont {D.}~\bibnamefont {Juncker}},\
  }\bibfield  {title} {\bibinfo {title} {Capillary microfluidics in
  microchannels: from microfluidic networks to capillaric circuits},\
  }\href@noop {} {\bibfield  {journal} {\bibinfo  {journal} {Lab Chip}\
  }\textbf {\bibinfo {volume} {28}},\ \bibinfo {pages} {2323} (\bibinfo {year}
  {2018})}\BibitemShut {NoStop}%
\bibitem [{\citenamefont {Murshed}\ and\ \citenamefont
  {de~Castro}(2017)}]{Murshed:RenewSustainErgRev2017}%
  \BibitemOpen
  \bibfield  {author} {\bibinfo {author} {\bibfnamefont {S.~S.}\ \bibnamefont
  {Murshed}}\ and\ \bibinfo {author} {\bibfnamefont {C.~N.}\ \bibnamefont
  {de~Castro}},\ }\bibfield  {title} {\bibinfo {title} {A critical review of
  traditional and emerging techniques and fluids for electronics cooling},\
  }\href@noop {} {\bibfield  {journal} {\bibinfo  {journal}
  {Renew.~Sust.~Energ.~Rev.}\ }\textbf {\bibinfo {volume} {78}},\ \bibinfo
  {pages} {821} (\bibinfo {year} {2017})}\BibitemShut {NoStop}%
\bibitem [{\citenamefont {Yang}\ \emph {et~al.}(2003)\citenamefont {Yang},
  \citenamefont {Lu}, \citenamefont {Kostiuk},\ and\ \citenamefont
  {Kwok}}]{Yang:JMicroMechEng2003}%
  \BibitemOpen
  \bibfield  {author} {\bibinfo {author} {\bibfnamefont {J.}~\bibnamefont
  {Yang}}, \bibinfo {author} {\bibfnamefont {F.}~\bibnamefont {Lu}}, \bibinfo
  {author} {\bibfnamefont {L.~W.}\ \bibnamefont {Kostiuk}},\ and\ \bibinfo
  {author} {\bibfnamefont {D.~Y.}\ \bibnamefont {Kwok}},\ }\bibfield  {title}
  {\bibinfo {title} {Electrokinetic microchannel battery by means of
  electrokinetic and microfluidic phenomena},\ }\href@noop {} {\bibfield
  {journal} {\bibinfo  {journal} {J.~Micromech. Microeng.}\ }\textbf {\bibinfo
  {volume} {13}},\ \bibinfo {pages} {963} (\bibinfo {year} {2003})}\BibitemShut
  {NoStop}%
\bibitem [{\citenamefont {van~der Heyden}\ \emph {et~al.}(2005)\citenamefont
  {van~der Heyden}, \citenamefont {Stein},\ and\ \citenamefont
  {Dekker}}]{vdHeyden:PRL2005}%
  \BibitemOpen
  \bibfield  {author} {\bibinfo {author} {\bibfnamefont {F.~H.~J.}\
  \bibnamefont {van~der Heyden}}, \bibinfo {author} {\bibfnamefont
  {D.}~\bibnamefont {Stein}},\ and\ \bibinfo {author} {\bibfnamefont
  {C.}~\bibnamefont {Dekker}},\ }\bibfield  {title} {\bibinfo {title}
  {Streaming currents in single nanofluidic channel},\ }\href@noop {}
  {\bibfield  {journal} {\bibinfo  {journal} {Phys. Rev. Lett.}\ }\textbf
  {\bibinfo {volume} {95}},\ \bibinfo {pages} {116104} (\bibinfo {year}
  {2005})}\BibitemShut {NoStop}%
\bibitem [{\citenamefont {Liu}\ \emph {et~al.}(2013)\citenamefont {Liu},
  \citenamefont {Hsu},\ and\ \citenamefont {Tseng}}]{Liu:Langmuir2013}%
  \BibitemOpen
  \bibfield  {author} {\bibinfo {author} {\bibfnamefont {K.~L.}\ \bibnamefont
  {Liu}}, \bibinfo {author} {\bibfnamefont {J.~P.}\ \bibnamefont {Hsu}},\ and\
  \bibinfo {author} {\bibfnamefont {S.}~\bibnamefont {Tseng}},\ }\bibfield
  {title} {\bibinfo {title} {Capillary osmosis in a charged nanopore connecting
  two large reservoirs},\ }\href@noop {} {\bibfield  {journal} {\bibinfo
  {journal} {Langmuir}\ }\textbf {\bibinfo {volume} {29}},\ \bibinfo {pages}
  {9598} (\bibinfo {year} {2013})}\BibitemShut {NoStop}%
\bibitem [{\citenamefont {Jing}\ and\ \citenamefont
  {Das}(2018)}]{Jing:PCCP2018}%
  \BibitemOpen
  \bibfield  {author} {\bibinfo {author} {\bibfnamefont {H.}~\bibnamefont
  {Jing}}\ and\ \bibinfo {author} {\bibfnamefont {S.}~\bibnamefont {Das}},\
  }\bibfield  {title} {\bibinfo {title} {Theory of diffusioosmosis in a charged
  nanochannel},\ }\href@noop {} {\bibfield  {journal} {\bibinfo  {journal}
  {Phys. Chem. Chem. Phys.}\ }\textbf {\bibinfo {volume} {20}},\ \bibinfo
  {pages} {10204} (\bibinfo {year} {2018})}\BibitemShut {NoStop}%
\bibitem [{\citenamefont {Bregulla}\ \emph {et~al.}(2016)\citenamefont
  {Bregulla}, \citenamefont {W\"urger}, \citenamefont {G\"unther},
  \citenamefont {Mertig},\ and\ \citenamefont {Cichos}}]{Bregulla:PRL2016}%
  \BibitemOpen
  \bibfield  {author} {\bibinfo {author} {\bibfnamefont {A.~P.}\ \bibnamefont
  {Bregulla}}, \bibinfo {author} {\bibfnamefont {A.}~\bibnamefont {W\"urger}},
  \bibinfo {author} {\bibfnamefont {K.}~\bibnamefont {G\"unther}}, \bibinfo
  {author} {\bibfnamefont {M.}~\bibnamefont {Mertig}},\ and\ \bibinfo {author}
  {\bibfnamefont {F.}~\bibnamefont {Cichos}},\ }\bibfield  {title} {\bibinfo
  {title} {Thermo-osmotic flow in thin films},\ }\href@noop {} {\bibfield
  {journal} {\bibinfo  {journal} {Phys.~Rev.~Lett.}\ }\textbf {\bibinfo
  {volume} {116}},\ \bibinfo {pages} {188303} (\bibinfo {year}
  {2016})}\BibitemShut {NoStop}%
\bibitem [{\citenamefont {Dietzel}\ and\ \citenamefont
  {Hardt}(2017)}]{Dietzel:JFM2017}%
  \BibitemOpen
  \bibfield  {author} {\bibinfo {author} {\bibfnamefont {M.}~\bibnamefont
  {Dietzel}}\ and\ \bibinfo {author} {\bibfnamefont {S.}~\bibnamefont
  {Hardt}},\ }\bibfield  {title} {\bibinfo {title} {Flow and streaming
  potential of an electrolyte in a channel with an axial temperature
  gradient},\ }\href@noop {} {\bibfield  {journal} {\bibinfo  {journal}
  {J.~Fluid Mech.}\ }\textbf {\bibinfo {volume} {813}},\ \bibinfo {pages}
  {1060} (\bibinfo {year} {2017})}\BibitemShut {NoStop}%
\bibitem [{\citenamefont {Rubinstein}\ and\ \citenamefont
  {Zaltzman}(2000)}]{Rubinstein:PRE2000}%
  \BibitemOpen
  \bibfield  {author} {\bibinfo {author} {\bibfnamefont {I.}~\bibnamefont
  {Rubinstein}}\ and\ \bibinfo {author} {\bibfnamefont {B.}~\bibnamefont
  {Zaltzman}},\ }\bibfield  {title} {\bibinfo {title} {Electro-osmotically
  induced convection at a permselective membrane},\ }\href@noop {} {\bibfield
  {journal} {\bibinfo  {journal} {Phys.~Rev.~E}\ }\textbf {\bibinfo {volume}
  {62}},\ \bibinfo {pages} {2238} (\bibinfo {year} {2000})}\BibitemShut
  {NoStop}%
\bibitem [{\citenamefont {Green}\ \emph {et~al.}(2000)\citenamefont {Green},
  \citenamefont {Ramos}, \citenamefont {Gonz\'alez}, \citenamefont {Morgan},\
  and\ \citenamefont {Castellanos}}]{Green:PRE2000}%
  \BibitemOpen
  \bibfield  {author} {\bibinfo {author} {\bibfnamefont {N.~G.}\ \bibnamefont
  {Green}}, \bibinfo {author} {\bibfnamefont {A.}~\bibnamefont {Ramos}},
  \bibinfo {author} {\bibfnamefont {A.}~\bibnamefont {Gonz\'alez}}, \bibinfo
  {author} {\bibfnamefont {H.}~\bibnamefont {Morgan}},\ and\ \bibinfo {author}
  {\bibfnamefont {A.}~\bibnamefont {Castellanos}},\ }\bibfield  {title}
  {\bibinfo {title} {Fluid flow induced by nonuniform ac electric fields in
  electrolytes on microelectrodes. i. experimental measurements},\ }\href@noop
  {} {\bibfield  {journal} {\bibinfo  {journal} {Phys.~Rev.~E}\ }\textbf
  {\bibinfo {volume} {61}},\ \bibinfo {pages} {4011} (\bibinfo {year}
  {2000})}\BibitemShut {NoStop}%
\bibitem [{\citenamefont {Gonz\'alez}\ \emph {et~al.}(2000)\citenamefont
  {Gonz\'alez}, \citenamefont {Ramos}, \citenamefont {Green}, \citenamefont
  {Castellanos},\ and\ \citenamefont {Morgan}}]{Gonzalez:PRE2000}%
  \BibitemOpen
  \bibfield  {author} {\bibinfo {author} {\bibfnamefont {A.}~\bibnamefont
  {Gonz\'alez}}, \bibinfo {author} {\bibfnamefont {A.}~\bibnamefont {Ramos}},
  \bibinfo {author} {\bibfnamefont {N.~G.}\ \bibnamefont {Green}}, \bibinfo
  {author} {\bibfnamefont {A.}~\bibnamefont {Castellanos}},\ and\ \bibinfo
  {author} {\bibfnamefont {H.}~\bibnamefont {Morgan}},\ }\bibfield  {title}
  {\bibinfo {title} {Fluid flow induced by nonuniform ac electric fields in
  electrolytes on microelectrodes. ii. a linear double-layer analysis},\
  }\href@noop {} {\bibfield  {journal} {\bibinfo  {journal} {Phys.~Rev.~E}\
  }\textbf {\bibinfo {volume} {61}},\ \bibinfo {pages} {4019} (\bibinfo {year}
  {2000})}\BibitemShut {NoStop}%
\bibitem [{\citenamefont {Ramos}\ \emph {et~al.}(2005)\citenamefont {Ramos},
  \citenamefont {Morgan}, \citenamefont {Green}, \citenamefont {Gonz\'{a}lez},\
  and\ \citenamefont {Castellanos}}]{Ramos:JAP2005}%
  \BibitemOpen
  \bibfield  {author} {\bibinfo {author} {\bibfnamefont {A.}~\bibnamefont
  {Ramos}}, \bibinfo {author} {\bibfnamefont {H.}~\bibnamefont {Morgan}},
  \bibinfo {author} {\bibfnamefont {N.~G.}\ \bibnamefont {Green}}, \bibinfo
  {author} {\bibfnamefont {A.}~\bibnamefont {Gonz\'{a}lez}},\ and\ \bibinfo
  {author} {\bibfnamefont {A.}~\bibnamefont {Castellanos}},\ }\bibfield
  {title} {\bibinfo {title} {Pumping of liquids with traveling-wave
  electroosmosis},\ }\href@noop {} {\bibfield  {journal} {\bibinfo  {journal}
  {J.~Appl.~Phys.}\ }\textbf {\bibinfo {volume} {97}},\ \bibinfo {pages}
  {084906} (\bibinfo {year} {2005})}\BibitemShut {NoStop}%
\bibitem [{\citenamefont {Ramos}\ \emph {et~al.}(2008)\citenamefont {Ramos},
  \citenamefont {Gonz\'{a}lez}, \citenamefont {Garc\'{i}a-S\'{a}nchez},\ and\
  \citenamefont {Castellanos}}]{Ramos:JCollIntScie2007}%
  \BibitemOpen
  \bibfield  {author} {\bibinfo {author} {\bibfnamefont {A.}~\bibnamefont
  {Ramos}}, \bibinfo {author} {\bibfnamefont {A.}~\bibnamefont {Gonz\'{a}lez}},
  \bibinfo {author} {\bibfnamefont {P.}~\bibnamefont
  {Garc\'{i}a-S\'{a}nchez}},\ and\ \bibinfo {author} {\bibfnamefont
  {A.}~\bibnamefont {Castellanos}},\ }\bibfield  {title} {\bibinfo {title} {A
  linear analysis of the effect of $\textrm{F}$aradaic currents on traveling
  wave electroosmosis},\ }\href@noop {} {\bibfield  {journal} {\bibinfo
  {journal} {J.~Colloid Interf.~Sci.}\ }\textbf {\bibinfo {volume} {309}},\
  \bibinfo {pages} {323} (\bibinfo {year} {2008})}\BibitemShut {NoStop}%
\bibitem [{\citenamefont {Gonz\'alez}\ \emph {et~al.}(2008)\citenamefont
  {Gonz\'alez}, \citenamefont {Ramos},\ and\ \citenamefont
  {Castellanos}}]{Gonzalez:MANO2008}%
  \BibitemOpen
  \bibfield  {author} {\bibinfo {author} {\bibfnamefont {A.}~\bibnamefont
  {Gonz\'alez}}, \bibinfo {author} {\bibfnamefont {A.}~\bibnamefont {Ramos}},\
  and\ \bibinfo {author} {\bibfnamefont {A.}~\bibnamefont {Castellanos}},\
  }\bibfield  {title} {\bibinfo {title} {Pumping of electrolytes using
  travelling-wave electro-osmosis: a weakly nonlinear analysis},\ }\href@noop
  {} {\bibfield  {journal} {\bibinfo  {journal} {Microfluid.~Nanofluid.}\
  }\textbf {\bibinfo {volume} {5}},\ \bibinfo {pages} {507} (\bibinfo {year}
  {2008})}\BibitemShut {NoStop}%
\bibitem [{\citenamefont {Yang}\ \emph {et~al.}(2009)\citenamefont {Yang},
  \citenamefont {Jiang}, \citenamefont {Shang}, \citenamefont {Ramos},\ and\
  \citenamefont
  {Garc\'{i}a-S\'{a}nchez}}]{Yang:IEEETransDielectricsElectricIns2009}%
  \BibitemOpen
  \bibfield  {author} {\bibinfo {author} {\bibfnamefont {H.}~\bibnamefont
  {Yang}}, \bibinfo {author} {\bibfnamefont {H.}~\bibnamefont {Jiang}},
  \bibinfo {author} {\bibfnamefont {D.}~\bibnamefont {Shang}}, \bibinfo
  {author} {\bibfnamefont {A.}~\bibnamefont {Ramos}},\ and\ \bibinfo {author}
  {\bibfnamefont {P.}~\bibnamefont {Garc\'{i}a-S\'{a}nchez}},\ }\bibfield
  {title} {\bibinfo {title} {Experiments on traveling-wave electroomosis:
  Effect of electrolyte conductivity},\ }\href@noop {} {\bibfield  {journal}
  {\bibinfo  {journal} {(IEEE) Trans. Dielectr. Electr. Insul.}\ }\textbf
  {\bibinfo {volume} {16}},\ \bibinfo {pages} {417} (\bibinfo {year}
  {2009})}\BibitemShut {NoStop}%
\bibitem [{\citenamefont {Yeh}\ \emph {et~al.}(2011)\citenamefont {Yeh},
  \citenamefont {Yang},\ and\ \citenamefont {Luo}}]{Yeh:PRE2011}%
  \BibitemOpen
  \bibfield  {author} {\bibinfo {author} {\bibfnamefont {H.~C.}\ \bibnamefont
  {Yeh}}, \bibinfo {author} {\bibfnamefont {R.~J.}\ \bibnamefont {Yang}},\ and\
  \bibinfo {author} {\bibfnamefont {W.~J.}\ \bibnamefont {Luo}},\ }\bibfield
  {title} {\bibinfo {title} {Analysis of traveling-wave electro-osmotic pumping
  with double-sided electrode arrays},\ }\href@noop {} {\bibfield  {journal}
  {\bibinfo  {journal} {Phys.~Rev.~E}\ }\textbf {\bibinfo {volume} {83}},\
  \bibinfo {pages} {056326} (\bibinfo {year} {2011})}\BibitemShut {NoStop}%
\bibitem [{\citenamefont {Hrdlicka}\ \emph {et~al.}(2014)\citenamefont
  {Hrdlicka}, \citenamefont {Patel},\ and\ \citenamefont
  {Snita}}]{Hrdlicka:Electrophoresis2014}%
  \BibitemOpen
  \bibfield  {author} {\bibinfo {author} {\bibfnamefont {J.}~\bibnamefont
  {Hrdlicka}}, \bibinfo {author} {\bibfnamefont {N.~S.}\ \bibnamefont
  {Patel}},\ and\ \bibinfo {author} {\bibfnamefont {D.}~\bibnamefont {Snita}},\
  }\bibfield  {title} {\bibinfo {title} {Traveling wave electroosmosis: The
  influence of electrode array geometry},\ }\href@noop {} {\bibfield  {journal}
  {\bibinfo  {journal} {Electrophoresis}\ }\textbf {\bibinfo {volume} {35}},\
  \bibinfo {pages} {1790} (\bibinfo {year} {2014})}\BibitemShut {NoStop}%
\bibitem [{\citenamefont {Kaatze}(1989)}]{Kaatze:JChemEngData1989}%
  \BibitemOpen
  \bibfield  {author} {\bibinfo {author} {\bibfnamefont {U.}~\bibnamefont
  {Kaatze}},\ }\bibfield  {title} {\bibinfo {title} {Complex permittivity of
  water as a function of frequency and temperature},\ }\href@noop {} {\bibfield
   {journal} {\bibinfo  {journal} {J.~Chem.~Eng.~Data}\ }\textbf {\bibinfo
  {volume} {34}},\ \bibinfo {pages} {371} (\bibinfo {year} {1989})}\BibitemShut
  {NoStop}%
\bibitem [{\citenamefont {Levich}(1962)}]{Levich:Prentice1962}%
  \BibitemOpen
  \bibfield  {author} {\bibinfo {author} {\bibfnamefont {V.~G.}\ \bibnamefont
  {Levich}},\ }in\ \href@noop {} {\emph {\bibinfo {booktitle} {Physicochemical
  Hydrodynamics}}}\ (\bibinfo  {publisher} {Prentice-Hall, New Jersey},\
  \bibinfo {year} {1962})\BibitemShut {NoStop}%
\bibitem [{\citenamefont {Pascall}\ and\ \citenamefont
  {Squires}(2011)}]{Pascall:JFM2011}%
  \BibitemOpen
  \bibfield  {author} {\bibinfo {author} {\bibfnamefont {A.~J.}\ \bibnamefont
  {Pascall}}\ and\ \bibinfo {author} {\bibfnamefont {T.~M.}\ \bibnamefont
  {Squires}},\ }\bibfield  {title} {\bibinfo {title} {Electrokinetics at
  liquid/liquid interfaces},\ }\href@noop {} {\bibfield  {journal} {\bibinfo
  {journal} {J.~Fluid Mech.}\ }\textbf {\bibinfo {volume} {684}},\ \bibinfo
  {pages} {163} (\bibinfo {year} {2011})}\BibitemShut {NoStop}%
\bibitem [{\citenamefont {Buchner}\ \emph {et~al.}(1999)\citenamefont
  {Buchner}, \citenamefont {Hefter},\ and\ \citenamefont
  {May}}]{Buchner:PhysChemA1999}%
  \BibitemOpen
  \bibfield  {author} {\bibinfo {author} {\bibfnamefont {R.}~\bibnamefont
  {Buchner}}, \bibinfo {author} {\bibfnamefont {G.~T.}\ \bibnamefont
  {Hefter}},\ and\ \bibinfo {author} {\bibfnamefont {P.~M.}\ \bibnamefont
  {May}},\ }\bibfield  {title} {\bibinfo {title} {Dielectric relaxation of
  aqueous $\textrm{NaCl}$ solutions},\ }\href@noop {} {\bibfield  {journal}
  {\bibinfo  {journal} {J.~ Phys.~Chem.~A}\ }\textbf {\bibinfo {volume}
  {103}},\ \bibinfo {pages} {1} (\bibinfo {year} {1999})}\BibitemShut {NoStop}%
\bibitem [{\citenamefont {Olesen}\ \emph {et~al.}(2010)\citenamefont {Olesen},
  \citenamefont {Bazant},\ and\ \citenamefont {Bruus}}]{Olesen:PRE2010}%
  \BibitemOpen
  \bibfield  {author} {\bibinfo {author} {\bibfnamefont {L.~H.}\ \bibnamefont
  {Olesen}}, \bibinfo {author} {\bibfnamefont {M.~Z.}\ \bibnamefont {Bazant}},\
  and\ \bibinfo {author} {\bibfnamefont {H.}~\bibnamefont {Bruus}},\ }\bibfield
   {title} {\bibinfo {title} {Strong nonlinear dynamics of electrolytes in
  large ac voltages},\ }\href@noop {} {\bibfield  {journal} {\bibinfo
  {journal} {Phys.~Rev.~E}\ }\textbf {\bibinfo {volume} {82}},\ \bibinfo
  {pages} {011501} (\bibinfo {year} {2010})}\BibitemShut {NoStop}%
\bibitem [{\citenamefont {Werkhoven}\ \emph {et~al.}(2018)\citenamefont
  {Werkhoven}, \citenamefont {Everts}, \citenamefont {Samin},\ and\
  \citenamefont {van Roij}}]{Werkhoven:PRL2018}%
  \BibitemOpen
  \bibfield  {author} {\bibinfo {author} {\bibfnamefont {B.~L.}\ \bibnamefont
  {Werkhoven}}, \bibinfo {author} {\bibfnamefont {J.~C.}\ \bibnamefont
  {Everts}}, \bibinfo {author} {\bibfnamefont {S.}~\bibnamefont {Samin}},\ and\
  \bibinfo {author} {\bibfnamefont {R.}~\bibnamefont {van Roij}},\ }\bibfield
  {title} {\bibinfo {title} {Flow-induced surface charge heterogeneity in
  electrokinetics due to $\textrm{S}$tern-layer conductance coupled to reaction
  kinetics},\ }\href@noop {} {\bibfield  {journal} {\bibinfo  {journal}
  {Phys.~Rev.~Lett.}\ }\textbf {\bibinfo {volume} {120}},\ \bibinfo {pages}
  {264502} (\bibinfo {year} {2018})}\BibitemShut {NoStop}%
\bibitem [{\citenamefont {Lei}\ \emph {et~al.}(2014)\citenamefont {Lei},
  \citenamefont {Hill},\ and\ \citenamefont {Glynne-Jones}}]{Lei:LabChip2014}%
  \BibitemOpen
  \bibfield  {author} {\bibinfo {author} {\bibfnamefont {J.}~\bibnamefont
  {Lei}}, \bibinfo {author} {\bibfnamefont {M.}~\bibnamefont {Hill}},\ and\
  \bibinfo {author} {\bibfnamefont {P.}~\bibnamefont {Glynne-Jones}},\
  }\bibfield  {title} {\bibinfo {title} {Numerical simulation of 3$\textrm{D}$
  boundary-driven acoustic streaming in microfluidic devices},\ }\href@noop {}
  {\bibfield  {journal} {\bibinfo  {journal} {Lab Chip}\ }\textbf {\bibinfo
  {volume} {14}},\ \bibinfo {pages} {532} (\bibinfo {year} {2014})}\BibitemShut
  {NoStop}%
\bibitem [{\citenamefont {Bazant}\ \emph {et~al.}(2009)\citenamefont {Bazant},
  \citenamefont {Kilic}, \citenamefont {Storey},\ and\ \citenamefont
  {Ajdari}}]{Bazant:AdvCollIntScie2009}%
  \BibitemOpen
  \bibfield  {author} {\bibinfo {author} {\bibfnamefont {M.~Z.}\ \bibnamefont
  {Bazant}}, \bibinfo {author} {\bibfnamefont {M.~S.}\ \bibnamefont {Kilic}},
  \bibinfo {author} {\bibfnamefont {B.~D.}\ \bibnamefont {Storey}},\ and\
  \bibinfo {author} {\bibfnamefont {A.}~\bibnamefont {Ajdari}},\ }\bibfield
  {title} {\bibinfo {title} {Towards an understanding of induced-charge
  electrokinetics at large applied voltages in concentrated solutions},\
  }\href@noop {} {\bibfield  {journal} {\bibinfo  {journal} {Adv.~Colloid
  Interfac.}\ }\textbf {\bibinfo {volume} {152}},\ \bibinfo {pages} {48}
  (\bibinfo {year} {2009})},\ \bibinfo {note} {and references
  therein}\BibitemShut {NoStop}%
\bibitem [{\citenamefont {Bazant}\ \emph {et~al.}(2011)\citenamefont {Bazant},
  \citenamefont {Storey},\ and\ \citenamefont {Kornyshev}}]{Bazant:PRL2011}%
  \BibitemOpen
  \bibfield  {author} {\bibinfo {author} {\bibfnamefont {M.~Z.}\ \bibnamefont
  {Bazant}}, \bibinfo {author} {\bibfnamefont {B.~D.}\ \bibnamefont {Storey}},\
  and\ \bibinfo {author} {\bibfnamefont {A.~A.}\ \bibnamefont {Kornyshev}},\
  }\bibfield  {title} {\bibinfo {title} {Double layer in ionic liquids:
  Overscreening versus crowding},\ }\href@noop {} {\bibfield  {journal}
  {\bibinfo  {journal} {Phys.~Rev.~Lett.}\ }\textbf {\bibinfo {volume} {106}},\
  \bibinfo {pages} {046102} (\bibinfo {year} {2011})}\BibitemShut {NoStop}%
\bibitem [{\citenamefont {Kilic}\ \emph
  {et~al.}(2007{\natexlab{a}})\citenamefont {Kilic}, \citenamefont {Bazant},\
  and\ \citenamefont {Ajdari}}]{Kilic:PRE2007b}%
  \BibitemOpen
  \bibfield  {author} {\bibinfo {author} {\bibfnamefont {M.~S.}\ \bibnamefont
  {Kilic}}, \bibinfo {author} {\bibfnamefont {M.~Z.}\ \bibnamefont {Bazant}},\
  and\ \bibinfo {author} {\bibfnamefont {A.}~\bibnamefont {Ajdari}},\
  }\bibfield  {title} {\bibinfo {title} {Steric effects in the dynamics of
  electrolytes at large applied voltages. ii. modified
  $\textrm{P}$oisson-$\textrm{N}$ernst-$\textrm{P}$lanck equations},\
  }\href@noop {} {\bibfield  {journal} {\bibinfo  {journal} {Phys.~Rev.~E}\
  }\textbf {\bibinfo {volume} {75}},\ \bibinfo {pages} {021503} (\bibinfo
  {year} {2007}{\natexlab{a}})}\BibitemShut {NoStop}%
\bibitem [{\citenamefont {Greberg}\ and\ \citenamefont
  {Kjellander}(1998)}]{Greberg:JChemPhys1998}%
  \BibitemOpen
  \bibfield  {author} {\bibinfo {author} {\bibfnamefont {H.}~\bibnamefont
  {Greberg}}\ and\ \bibinfo {author} {\bibfnamefont {R.}~\bibnamefont
  {Kjellander}},\ }\bibfield  {title} {\bibinfo {title} {Charge inversion in
  electric double layers and effects of different sizes for counterions and
  coions},\ }\href@noop {} {\bibfield  {journal} {\bibinfo  {journal}
  {J.~Chem.~Phys.}\ }\textbf {\bibinfo {volume} {108}},\ \bibinfo {pages}
  {2940} (\bibinfo {year} {1998})}\BibitemShut {NoStop}%
\bibitem [{\citenamefont {Kilic}\ \emph
  {et~al.}(2007{\natexlab{b}})\citenamefont {Kilic}, \citenamefont {Bazant},\
  and\ \citenamefont {Ajdari}}]{Kilic:PRE2007a}%
  \BibitemOpen
  \bibfield  {author} {\bibinfo {author} {\bibfnamefont {M.~S.}\ \bibnamefont
  {Kilic}}, \bibinfo {author} {\bibfnamefont {M.~Z.}\ \bibnamefont {Bazant}},\
  and\ \bibinfo {author} {\bibfnamefont {A.}~\bibnamefont {Ajdari}},\
  }\bibfield  {title} {\bibinfo {title} {Steric effects in the dynamics of
  electrolytes at large applied voltages. i. double layer charging},\
  }\href@noop {} {\bibfield  {journal} {\bibinfo  {journal} {Phys.~Rev.~E}\
  }\textbf {\bibinfo {volume} {75}},\ \bibinfo {pages} {021502} (\bibinfo
  {year} {2007}{\natexlab{b}})}\BibitemShut {NoStop}%
\bibitem [{Com()}]{Comsol2019}%
  \BibitemOpen
  \href@noop {} {}\bibinfo {note} {Comsol Multiphysics, COMSOL
  \textsuperscript{\textregistered}, Inc., G\"ottingen, Germany,
  2019}\BibitemShut {NoStop}%
\bibitem [{\citenamefont {Suh}\ and\ \citenamefont
  {Kang}(2011)}]{Suh:IJNumMethFluids2011}%
  \BibitemOpen
  \bibfield  {author} {\bibinfo {author} {\bibfnamefont {Y.~K.}\ \bibnamefont
  {Suh}}\ and\ \bibinfo {author} {\bibfnamefont {S.}~\bibnamefont {Kang}},\
  }\bibfield  {title} {\bibinfo {title} {Simple, coupled algorithms for solving
  creeping flows and their application to electro-osmotic flows},\ }\href@noop
  {} {\bibfield  {journal} {\bibinfo  {journal} {Int.~J.~Numer.~Meth.~Fluids}\
  }\textbf {\bibinfo {volume} {66}},\ \bibinfo {pages} {1248} (\bibinfo {year}
  {2011})}\BibitemShut {NoStop}%
\bibitem [{\citenamefont {Suh}(2011)}]{Suh:CollSurfA2011}%
  \BibitemOpen
  \bibfield  {author} {\bibinfo {author} {\bibfnamefont {Y.~K.}\ \bibnamefont
  {Suh}},\ }\bibfield  {title} {\bibinfo {title} {Numerical study on transient
  induced-charge electro-osmotic flow in a cavity},\ }\href@noop {} {\bibfield
  {journal} {\bibinfo  {journal} {Colloids Surf.~A}\ }\textbf {\bibinfo
  {volume} {376}},\ \bibinfo {pages} {111} (\bibinfo {year}
  {2011})}\BibitemShut {NoStop}%
\bibitem [{\citenamefont {Green}\ \emph {et~al.}(2002)\citenamefont {Green},
  \citenamefont {Ramos}, \citenamefont {Gonz\'alez}, \citenamefont {Morgan},\
  and\ \citenamefont {Castellanos}}]{Green:PRE2002}%
  \BibitemOpen
  \bibfield  {author} {\bibinfo {author} {\bibfnamefont {N.~G.}\ \bibnamefont
  {Green}}, \bibinfo {author} {\bibfnamefont {A.}~\bibnamefont {Ramos}},
  \bibinfo {author} {\bibfnamefont {A.}~\bibnamefont {Gonz\'alez}}, \bibinfo
  {author} {\bibfnamefont {H.}~\bibnamefont {Morgan}},\ and\ \bibinfo {author}
  {\bibfnamefont {A.}~\bibnamefont {Castellanos}},\ }\bibfield  {title}
  {\bibinfo {title} {Fluid flow induced by nonuniform $\textrm{AC}$ electric
  fields in electrolytes on microelectrodes. iii. observation of streamlines
  and numerical simulation},\ }\href@noop {} {\bibfield  {journal} {\bibinfo
  {journal} {Phys.~Rev.~E}\ }\textbf {\bibinfo {volume} {66}},\ \bibinfo
  {pages} {026305} (\bibinfo {year} {2002})}\BibitemShut {NoStop}%
\bibitem [{\citenamefont {Pribyl}\ \emph {et~al.}(2008)\citenamefont {Pribyl},
  \citenamefont {Snita},\ and\ \citenamefont {Marek}}]{Pribyl:InTech2008}%
  \BibitemOpen
  \bibfield  {author} {\bibinfo {author} {\bibfnamefont {M.}~\bibnamefont
  {Pribyl}}, \bibinfo {author} {\bibfnamefont {D.}~\bibnamefont {Snita}},\ and\
  \bibinfo {author} {\bibfnamefont {M.}~\bibnamefont {Marek}},\ }\bibfield
  {title} {\bibinfo {title} {Multiphysical modeling of $\textrm{DC}$ and
  $\textrm{AC}$ electroosmosis in micro- and nanosystems},\ }in\ \href
  {https://doi.org/10.5772/5969} {\emph {\bibinfo {booktitle} {Modeling and
  Simulation}}},\ \bibinfo {editor} {edited by\ \bibinfo {editor}
  {\bibfnamefont {G.}~\bibnamefont {Petrone}}\ and\ \bibinfo {editor}
  {\bibfnamefont {G.}~\bibnamefont {Cammarata}}}\ (\bibinfo  {publisher}
  {InTechOpen},\ \bibinfo {year} {2008})\BibitemShut {NoStop}%
\bibitem [{\citenamefont {Pribyl}\ \emph {et~al.}(2006)\citenamefont {Pribyl},
  \citenamefont {Schrott}, \citenamefont {Shahravan},\ and\ \citenamefont
  {Sinta}}]{Pribyl:ISEHD2006}%
  \BibitemOpen
  \bibfield  {author} {\bibinfo {author} {\bibfnamefont {M.}~\bibnamefont
  {Pribyl}}, \bibinfo {author} {\bibfnamefont {W.}~\bibnamefont {Schrott}},
  \bibinfo {author} {\bibfnamefont {A.}~\bibnamefont {Shahravan}},\ and\
  \bibinfo {author} {\bibfnamefont {D.}~\bibnamefont {Sinta}},\ }\bibfield
  {title} {\bibinfo {title} {Mathematical modeling of a microchip driven by
  $\textrm{AC}$ electric field},\ }in\ \href@noop {} {\emph {\bibinfo
  {booktitle} {Proceedings of 2006 IEEE International Symposium on
  Electrohydrodynamics}}}\ (\bibinfo {year} {2006})\ pp.\ \bibinfo {pages}
  {309--312}\BibitemShut {NoStop}%
\bibitem [{\citenamefont {Sugioka}(2016)}]{Sugioka:PRE2016}%
  \BibitemOpen
  \bibfield  {author} {\bibinfo {author} {\bibfnamefont {H.}~\bibnamefont
  {Sugioka}},\ }\bibfield  {title} {\bibinfo {title} {Direct simulation of
  phase delay effects on induced-charge electro-osmosis under large ac electric
  fields},\ }\href@noop {} {\bibfield  {journal} {\bibinfo  {journal}
  {Phys.~Rev.~E}\ }\textbf {\bibinfo {volume} {94}},\ \bibinfo {pages} {022609}
  (\bibinfo {year} {2016})}\BibitemShut {NoStop}%
\bibitem [{\citenamefont {Gonz\'alez}\ \emph {et~al.}(2010)\citenamefont
  {Gonz\'alez}, \citenamefont {Ramos}, \citenamefont {Garc\'ia-S\'anchez},\
  and\ \citenamefont {Castellanos}}]{Gonzalez:PRE2010}%
  \BibitemOpen
  \bibfield  {author} {\bibinfo {author} {\bibfnamefont {A.}~\bibnamefont
  {Gonz\'alez}}, \bibinfo {author} {\bibfnamefont {A.}~\bibnamefont {Ramos}},
  \bibinfo {author} {\bibfnamefont {P.}~\bibnamefont {Garc\'ia-S\'anchez}},\
  and\ \bibinfo {author} {\bibfnamefont {A.}~\bibnamefont {Castellanos}},\
  }\bibfield  {title} {\bibinfo {title} {Effect of the combined action of
  $\textrm{F}$aradaic currents and mobility differences in ac
  electro-osmosis},\ }\href@noop {} {\bibfield  {journal} {\bibinfo  {journal}
  {Phys.~Rev.~E}\ }\textbf {\bibinfo {volume} {81}},\ \bibinfo {pages} {016320}
  (\bibinfo {year} {2010})}\BibitemShut {NoStop}%
\end{thebibliography}%

\end{document}